# AI Decodes Historical Chinese Archives to Reveal Lost Climate History

**Authors:** Sida He[1], Lingxi Xie[1], Xiaopeng Zhang[1], Qi Tian[1]*

**Affiliations:**

[1]Huawei Technologies Co., Ltd.; Shenzhen, China.

*Corresponding author. Email: tian.qi1@huawei.com

**Abstract:** Historical archives contain qualitative descriptions of climate events, yet converting these into quantitative records has remained a fundamental challenge. Here we introduce a paradigm shift: a generative AI framework that inverts the logic of historical chroniclers by inferring the quantitative climate patterns associated with documented events. Applied to historical Chinese archives, it produces the sub-annual precipitation reconstruction for southeastern China over the period 1368–1911 AD. Our reconstruction not only quantifies iconic extremes like the Ming Dynasty's Great Drought but also, crucially, maps the full spatial and seasonal structure of El Niño influence on precipitation in this region over five centuries, revealing dynamics inaccessible in shorter modern records. Our methodology and high-resolution climate dataset are directly applicable to climate science and have broader implications for the historical and social sciences.

Reconstructing the historical climate is crucial for understanding past societal and environmental changes (*1–3*), and for improving the fidelity of climate models (*4–6*). Historical climate reconstruction relies on proxies. Physical proxies (e.g., tree rings, ice cores) often lack sub-annual resolution. Historical texts, particularly from densely populated regions with comprehensive archives, can provide higher (annual or better) resolution records. However, their qualitative and descriptive nature makes them inherently difficult to integrate into standard quantitative reconstruction frameworks.

Recent advances in artificial intelligence (AI) offer a promising solution. AI models excel at learning complex patterns from large datasets, as shown by many AI weather forecasting models (*7–9*). Critically, AI can also synthesize new, coherent data that conform to learned patterns (*10–12*). We therefore propose a generative model which was trained on the growing volume of modern quantitative climate data (from observations and/or simulations), and can learn to invert the historical chronicler's logic: it can infer the quantitative climate patterns that most likely gave rise to the qualitative events documented in texts.

In this work, we focus on China, leveraging its unparalleled historical documentary tradition. We trained a diffusion model to learn the relationship between event sequences and precipitation fields. For the first time, we achieve a seasonal-to-monthly resolution reconstruction of precipitation in the populated region of southeastern China (bounded by the Heihe–Tengchong Line to the west, the 41°N parallel to the north, and all islands excluded; see Fig. 1A) over the 544-year Ming–Qing period (1368–1911). Figure 1B provides a schematic overview of the end-to-end reconstruction framework. Validated through skill testing, cross-comparison, and consistency checks with present-day climatology, the reconstruction captures historical extremes and reveals the structure of large-scale climate dynamics on centennial timescales inaccessible to modern records. This work demonstrates a scalable, AI-driven paradigm for reconstructing historical climate from qualitative texts. The approach, which leverages generative AI, can be extended to other regions and periods globally, contingent on ongoing international collaboration in excavating and organizing new textual archives.

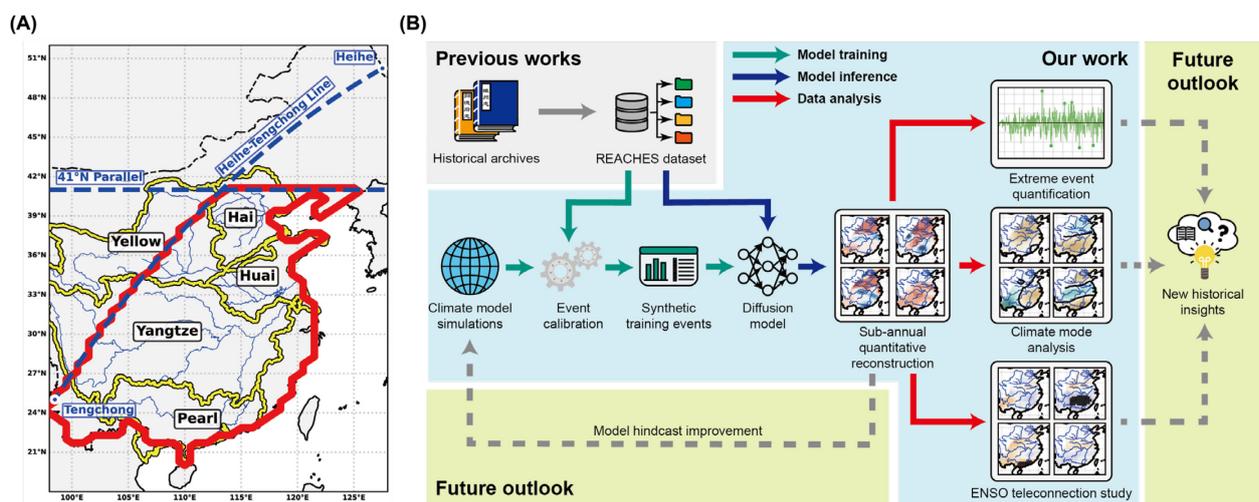

**Fig. 1. Generative AI paradigm for quantitative climate reconstruction from historical texts.** (A) Map of the study area. The analysis domain (within the red frame) is defined as the part of China bounded by the Heihe–Tengchong Line (*13*) to the west and the 41°N parallel to the north, with all islands excluded. This region contained ~84% of China's population in 2020 (*14*), matching high historical document density critical for text-based climate reconstruction. Major river basins analyzed (Hai, Yellow, Huai, Yangtze, and Pearl) are outlined (*15*). Note that portions of the Hai, Yellow, and Yangtze River basins falling

outside our study area were excluded from this analysis. (B) Schematic of the reconstruction framework. Our work builds upon the previous work of digitalizing the historical archives into the REACHES dataset (*16*). A 3D diffusion model is trained on pairs of synthetic climate events and corresponding precipitation fields generated from the climate model simulations. The event synthesis is calibrated to match the spatiotemporal density of documented events in the historical REACHES dataset. The trained model is then applied to actual historical climate events to generate quantitative, sub-annual precipitation reconstructions. These reconstructions enable downstream analyses (e.g., extreme event quantification, climate mode analysis, climate teleconnection studies) and provide a new resource for historical insights and climate model hindcasting.

**From texts to data**

Training our model relies on a corpus of paired quantitative precipitation data and corresponding climate records, enabling subsequent inference on historical textual archives. Three key challenges emerge: (A) sourcing reliable historical climate records, (B) assembling a sufficient corpus of quantitative precipitation data for training, and (C) establishing a robust linkage between precipitation magnitudes and their corresponding climate events.

**(A) Historical climate records.** We used the Reconstructed East Asian Climate Historical Encoded Series (REACHES) dataset (*16*), a digitized compilation of climate records spanning 1368–1911. The core unit of these records is a climate event, defined by its spatiotemporal location and type (e.g., flood, drought). This study focuses on precipitation; for this purpose, we defined four specific event categories—annual floods, annual droughts, sub-annual floods, and sub-annual droughts—and mapped the events from the REACHES dataset onto this taxonomy (see Materials and Methods Section 1.1 for details). Statistics on the number of events can be found in Supplementary Materials Figs. S2 and S3.

**(B) Quantitative precipitation data for training.** A generative AI model must first learn to reconstruct climate fields from climate events in a controlled setting. This requires abundant paired data of precipitation fields and their corresponding extreme events, far surpassing the volume of available observational datasets. We therefore turned to simulated data and selected the Community Earth System Model version 1 with the Community Atmospheric Model version 5 (CESM1-CAM5) (*17*, *18*) for the following reasons: (a) its relatively high spatial resolution (~1°) compared to its alternatives; (b) its large multi-scenario ensemble provides the necessary volume of training data; (c) its demonstrated skill in simulating East Asian climate variability (*19*, *20*), particularly precipitation extremes (*21*), which is essential for our event-based reconstruction. A total of 1,710 years of simulated data was used for model training, with an additional 161 years reserved for validation. The data preprocessing pipeline is elaborated upon in Materials and Methods Sections 2.1 and 2.2. The robustness of our conclusions to this choice is confirmed by an ablation study using a different climate model as training data (see Supplementary Text Section 9.1). Future incorporation of even more physical simulations or observationally constrained reanalysis products can refine the results within this framework.

**(C) Linking precipitation to recorded events.** To train the model, we translated the quantitative precipitation simulations into discrete events, emulating what a historical chronicler would have recorded. We adopted the operational hypothesis that surviving records predominantly capture the most extreme events as perceived within a contemporary time window. We approximated the length of the time window as 31 years, including the target year, 15 years preceding it, and 15 years following it, aligning with the standard 30-year climatological baseline used in modern climate science for defining anomalies (*22*). Guided by this hypothesis, we calibrated our translation algorithm such that the total number of the most extreme events identified in the simulated data matched the number of events recorded in the historical documents. Further details of this calibration process are provided in Materials and Methods Section 2.3.

**A diffusion model for historical climate reconstruction**

We adopted a diffusion model for its inherent robustness to noisy data and natural suitability for uncertainty quantification (*11*, *23*). More fundamentally, our approach constitutes a paradigm shift in paleoclimate reconstruction: Whereas the traditional point-to-point regression-based (PPR-based) methods rely on prespecified mapping functions (*24*, *25*), our generative model directly learns the complex joint spatiotemporal probability distribution relating sequences of historical events to precipitation fields. This enables it to infer quantitative precipitation patterns from sparse and incomplete event records in a manner beyond the reach of conventional regression-based techniques.

Our implementation treats the two spatial dimensions and the temporal dimension jointly within a 3D diffusion framework. The model takes all annual and sub-annual event data for a given year as input and generates precipitation values for all the 12 months in a single inference step. Seasonal and annual precipitation totals are then calculated by summing the corresponding monthly values. This unified modeling framework eliminates the need to train separate models for different temporal scales and ensures consistency across them.

The model was trained from scratch using a U-Net backbone (architecture illustrated in Fig. S7) (*10*, *11*). Critically, we used a periodic conditioning prompt to condition the model on the start date of the input data series, enabling both data augmentation and proper accounting for the variable timing of the Chinese New Year within the Gregorian calendar. The model contains approximately 20 million parameters. Detailed model configuration information is provided in Materials and Methods Section 3.1.

**Model validation strategy**

Ideally, a historical climate reconstruction would be validated against coeval instrumental data (*6*, *24*). This is not feasible for our study for two fundamental reasons. First, reliable instrumental precipitation data are unavailable for the Ming–Qing era. Second, the advent of instrumental observations shifted event-recording paradigms from qualitative descriptions (e.g., "severe drought") to quantitative metrics (e.g., rainfall in mm), creating an inherent misalignment between historical and modern event records that prevents applying our method to recent periods. Therefore, we employed a three-step framework: (A) quantifying model skill on a 161-year validation dataset, with results detailed in Supplementary Text Section 6 and Table S6; (B) cross-validating against published reconstructions, with the comparison shown in Supplementary Text Section 7 and Fig. S8; and (C) checking physical consistency with modern climatology, which is integrated into the analyses of Empirical Orthogonal Function (EOF) and El Niño–Southern Oscillation (ENSO) teleconnections in the following sections. This comprehensive, multi-pronged approach provides robust, circumstantial evidence for the reconstruction's validity, overcoming the inherent validation challenges in historical climate research. As quantified in the Supplementary Text Section 6, reconstruction uncertainty is inherently higher for winter and at the monthly scale, which stems from the intrinsic sparsity of documentary records at these timescales. This highlights the potential for future improvements as more detailed historical records, especially for winter, become available through ongoing archival research.

**Quantifying historic catastrophes**

As a demonstration, we studied three years with extreme climate anomalies that impacted northern China during the Ming and Qing dynasties: 1593, 1640, and 1877, respectively. The historical context of these years is well-documented. The year 1593 was marked by severe flooding in the Huai River Basin (Fig. 1A). The rainy season, lasting from April to September, caused widespread crop failure and a devastating famine (*26*). The year 1640 represents the peak of the Chongzhen Drought (1637–1643), which is widely considered a key factor in the fall of the Ming Dynasty (*27*, *28*). In 1877, northern China experienced the most severe phase of the Guangxu Drought (1875–1879), which formed part of the severe drought affecting monsoon Asia (*24*, *29*) and resulted in over 10 million fatalities in northern China (*30*, *31*).

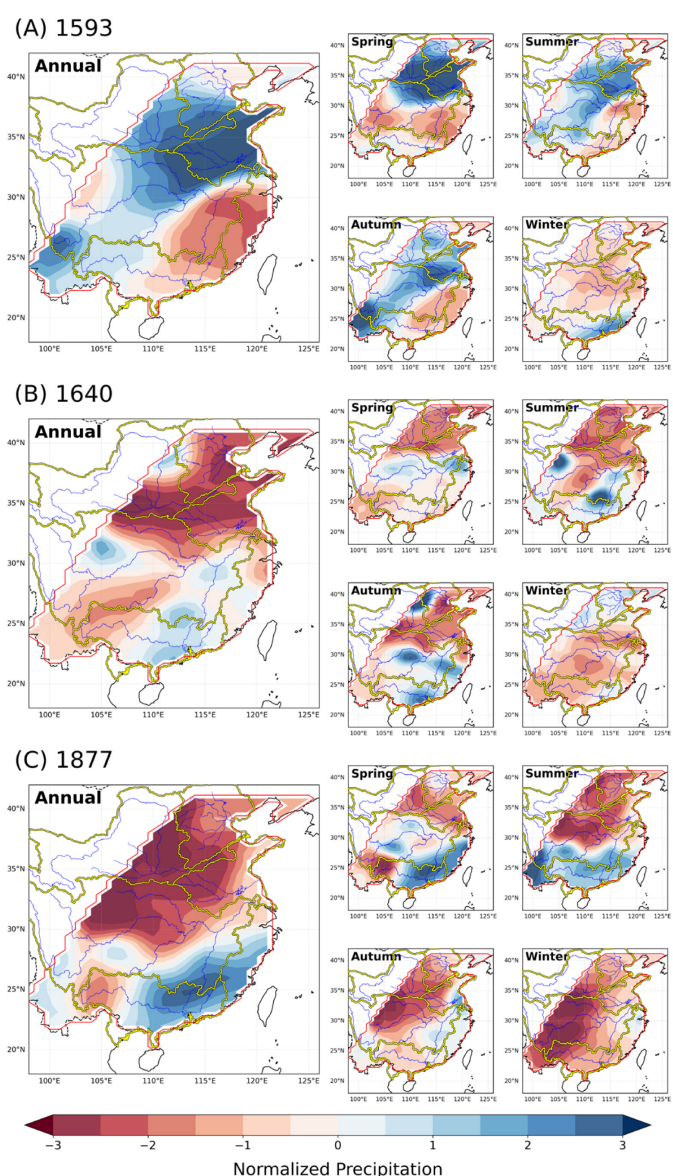

**Fig. 2. Reconstruction of iconic northern China precipitation extreme events.** Maps show reconstructed normalized annual and seasonal precipitation z-scores for (A) the great flood

of 1593, (B) the severe drought of 1640, and (C) the severe drought of 1877. The reconstruction uncovers the fine-scale spatial heterogeneity and seasonal development of these events, which were previously documented only qualitatively in historical texts. For example, it distinguishes the core flooding region in the Huai River Basin in 1593, the pervasive drought across northern China in 1640, and the heterogeneous drought pattern in 1877 with severe impacts concentrated in the west.

To quantify these events, we reconstructed the annual and seasonal precipitation z-scores, which represent the number of standard deviations a value deviates from the 31-year sliding window climatological mean, thus facilitating the comparison of anomalies across regions and seasons. The results are shown in Fig. 2. Our reconstruction clearly shows that in 1593, the core flooding area was centered in the Huai River Basin with an annual precipitation z-score exceeding +3. Above-average precipitation persisted through spring, summer, and autumn, with the most severe anomaly occurring in spring. In 1640, exceptional drought affected most of northern China, where the annual precipitation z-score was generally below −2 and persisted from spring to autumn. While the southern boundary of the 1640 drought approximated the Huai River, the 1877 drought extended further south to the Yangtze River. In northern China, the western part experienced extreme drought, whereas the eastern areas were less severely affected. Seasonal precipitation migration maps reveal that these eastern regions experienced only moderate drought in summer and even slightly wet conditions in autumn. A similar spatiotemporal pattern is also observed in Fig. 2 of (*32*), whose reconstruction of seasonal precipitation relied on an independent document that is not publicly available and not included in our database. This alignment excludes winter, consistent with our validation results indicating weaker model performance for winter precipitation. Overall, our reconstructions clearly illustrate the temporal evolution of these extreme climate events.

A key advantage of a continuous, quantitative reconstruction is that it enables a systematic comparison of extreme events across different historical periods, allowing us to objectively assess their relative severity and historical significance across centuries. For quantitative assessment, we calculated the average annual and seasonal precipitation z-scores for these three years across five major Chinese river basins: the Hai, Yellow, Huai, Yangtze, and Pearl, whose boundaries are delineated in Fig. 1A (*15*). We also ranked the drought and flood intensities for these years against the full 544-year Ming–Qing reconstruction. Results are provided in Tables S8–S10. As shown in Table S8, the annual precipitation z-score in 1593 ranked as the highest on record for the Huai River Basin and the third for the Yellow River Basin. Furthermore, precipitation in the Huai River Basin during spring, summer, and autumn each ranked among the wettest on record for their respective seasons. As shown in Table S9, the 1640 drought ranked second in severity for the Huai River Basin, and summer 1640 ranked among the three driest years on record for the Hai and Yellow River Basins. As shown in Table S10, in addition to confirming the exceptional drought in northern China in 1877, our reconstruction reveals that winter 1877 also marked a historical drought for the Yangtze River Basin, a region typically classified as part of southern China. Our ensemble-based reconstruction provides uncertainty estimates for these basin averages, quantified as the standard deviation across ensemble members (Table S7). For example, the 1593 flood in the Huai River Basin has an ensemble standard deviation of 0.12, and the 1877 drought in the Yellow River Basin has a standard deviation of 0.07. We conclude that such quantification facilitates direct comparisons and helps to contextualize the historical significance of the past extreme climate events.

This reconstruction also provides data at native monthly resolution, capable of revealing sub-seasonal climate dynamics such as the migration of precipitation anomalies during the 1721 drought (Supplementary Text Section 8 and Fig. S17). The elevated uncertainty of these monthly estimates, as quantified in our validation, should be considered in quantitative applications.

**Hidden patterns of historical climate**

Droughts and floods in China often exhibit specific spatial patterns, driven by the monsoon, the Western Pacific Subtropical High, and the Meiyu Front (*33*, *34*). To better characterize them in our reconstruction, we performed an EOF analysis on our annual precipitation dataset, as shown in Fig. 3A. The first four leading modes reveal distinct spatial structures: a uniform mode (Mode 1), a meridional dipolar mode (Mode 2), a zonal dipolar mode (Mode 3), and a meridional tripolar mode (Mode 4). The meridional dipolar and tripolar modes align with well-established features of East Asian monsoon precipitation variability, as extensively documented in previous studies (*35–38*). The uniform mode (Mode 1) is also a recurrent feature in long-term paleoclimate reconstructions (*39–42*) and has been physically linked to low-frequency variations in South China Sea surface temperatures, potentially exciting a large-scale, Gill-type atmospheric response (*39*). The zonal dipolar mode (Mode 3) highlights a fundamental east-west precipitation dipole, which reflects the differential influences of two major monsoon systems: the South Asian monsoon, which strongly influences southwestern China, and the East Asian monsoon, which dominates southeastern China (*43*, *44*). This mode also appears as Mode 4 from the CESM1-CAM5 training data (Fig. 3C). For a direct comparison with (*25*), we also performed an EOF analysis on May–September precipitation (Fig. S9). The resulting leading three modes, namely a uniform mode, a meridional dipolar mode, and a meridional tripolar mode, align with their reported patterns, as detailed in the Supplementary Text Section 7.

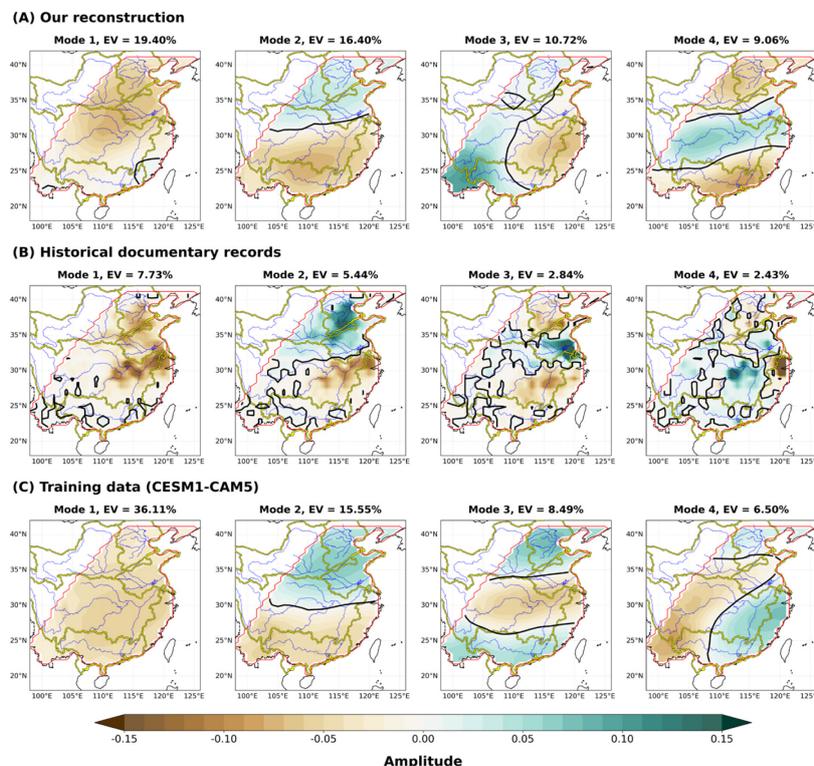

**Fig. 3. EOF analysis of climate variability.** Spatial patterns and explained variance (EV) for (A) our reconstruction, (B) gridded historical events, and (C) the training data (CESM1-CAM5 simulation). Our reconstruction synthesizes information from both the historical events and training data. For example, the region of maximum variability in Mode 1 (the uniform 'all-dry/all-wet' pattern) in our reconstruction is located at northern China, close to the historical events, while the spatial patterns of Modes 3 and 4 in our reconstruction are close to Modes 4 and 3 in the training data, respectively. Also, its variance concentration (EV values) is intermediate between the documentary and training data.

A key question is how our EOF patterns compare quantitatively with those derived from the raw documentary records and the climate model training data. To answer this, we also performed EOF analysis on both the historical documentary records and the training dataset. We quantified the documentary records into a gridded ternary dataset, matching the model input format (see Materials and Methods Section 3.2). The resulting modes are shown in Figs. 3B and 3C for the documentary records and training data, respectively. We find that Mode 1 exhibits a uniform pattern across our reconstruction, documentary records, and training data. The region of maximum variability for this mode in our reconstruction lies in northern China, aligning with documentary records. Similarly, Mode 2 is a meridional dipolar mode across all three datasets. Modes 3 and 4 of the training data display a meridional tripolar and a zonal dipolar pattern, respectively, matching Modes 4 and 3 of our reconstruction. A comparison of the explained variance (EV) for the leading modes reveals that the variance concentration of the reconstruction is intermediate between that of the documentary records with EV widely dispersed across modes and that of the training data with EV strongly concentrated in leading modes. This EOF analysis confirms that our reconstruction integrates information and preserves large-scale climate dynamics from both the historical documents and training data.

**Correlation with ENSO over the past half-millennium**

To investigate the response of the historical Chinese climate to ENSO, we analyzed the correlation between our reconstructed precipitation and the Niño 3.4 index from (*45*). Correlations for individual centuries (15th–19th) and the full Ming–Qing reconstruction are shown in Figs. 4A–4F. For a comparative analysis with the modern period, Fig. 4G presents correlations derived from the ERA5 reanalysis (1940–2024) (*46*) using a consistent Niño 3.4 index extended from (*45*) (see Materials and Methods Section 5).

As shown in Fig. 4F, our reconstruction for the annual precipitation shows a significant positive correlation with ENSO in southeastern China and a significant negative correlation in northern China, consistent with previous work (*25*, Fig. 9). Additionally, a significant negative correlation zone is identified in southwestern China, a feature that is present, though less pronounced, in both (*25*) and the reanalysis products in Fig. 4G. For seasonal precipitation, the correlation in southeastern China transitions from positive in spring to negative in winter, a pattern that aligns precisely with the ERA5 reanalysis.

Century-scale subsets (Figs. 4A-E) reveal variations in the strength and seasonal phasing of these correlations. The 19th-century pattern bears the closest resemblance to the modern ERA5 pattern. These apparent variations across earlier centuries illustrate our dataset's capacity to capture century-scale differences. The robust reconstruction of the teleconnection's core spatial and seasonal structure over the full five centuries establishes the essential baseline required to investigate genuine centennial-scale climate dynamics, a task impossible with the short instrumental record alone.

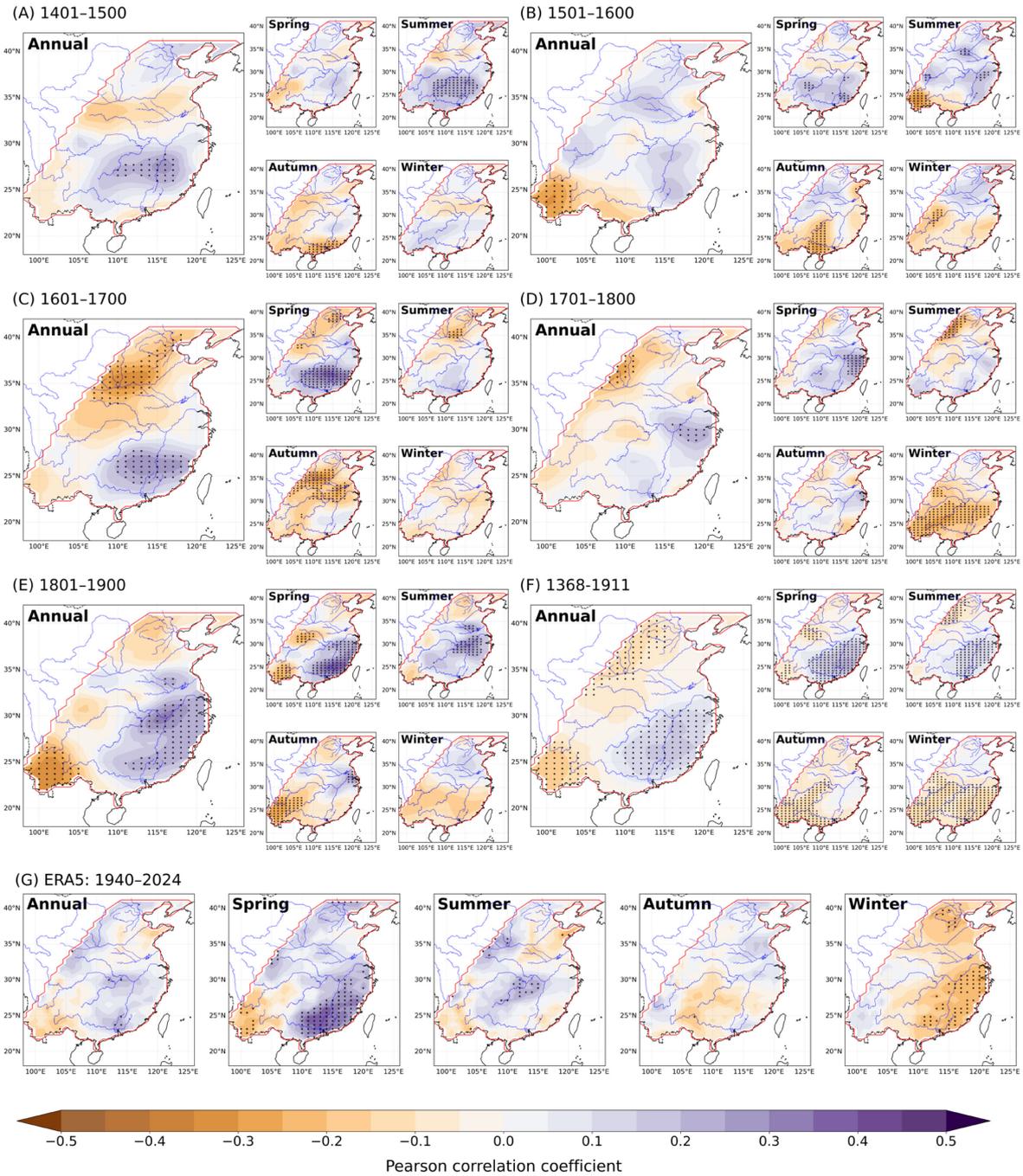

**Fig. 4. Evolution of the ENSO-precipitation correlation over centuries.** Panels (A–E) show the spatial correlation between the Niño 3.4 index and reconstructed annual and seasonal precipitation for individual centuries (15th–19th). Panel (F) shows the correlation for the full Ming–Qing reconstruction (1368–1911), and panel (G) shows the correlation for the modern ERA5 reanalysis (1940–2024). Stippling denotes statistical significance ($p < 0.05$). Our multi-century reconstruction (F) confirms the canonical annual dipole pattern (positive in southeastern China, negative in the north), in agreement with previous work, and reveals a statistically significant negative correlation in southwestern China. A seasonal phase reversal, positive in spring to negative in winter, in southeastern China is evident in both our reconstruction and the ERA5 reanalysis. By comparing Panels (A–E) to Panel (G), one can see that the 19th-century pattern most closely resembles the modern ERA5 pattern. Apparent variations across earlier centuries illustrate the dataset's capacity to capture century-scale

differences, underscoring the need for this extended quantitative record for investigating centennial-scale climate dynamics.

**Conclusions: A new frontier for paleoclimate and history**

We have established a novel generative AI paradigm to transform historical documents into quantitative climate data. Our key achievement is the first sub-annual precipitation reconstruction for southeastern China over 544 years (1368–1911). This paradigm offers fundamental methodological and practical advantages: it is highly efficient and reproducible, requiring minimal human intervention once trained, and it provides an objective framework that minimizes the subjectivity inherent in traditional qualitative archive analysis. The model shows competitive skill at the annual scale and provides robust uncertainty quantification. The reconstruction captures iconic extremes and, through EOF analysis, proves to be a skillful synthesis of large-scale climate patterns from simulations and historical records. Furthermore, our reconstruction establishes a continuous, quantitative record of the ENSO–East Asian monsoon teleconnection over the 544 years, providing an essential baseline for detecting and understanding genuine low-frequency variability that cannot be evaluated with the short instrumental record. To address the absence of instrumental benchmarks, our framework provides a transparent methodology validated through skill testing, cross-comparison, and physical consistency checks, building a multifaceted case for credibility that moves beyond reliance on a single validation source.

Despite these advances, several limitations and opportunities for future work remain. First, our current event-based approach, which utilizes a limited set of binary (occurrence/non-occurrence) categories, represents a significant under-utilization of the rich information contained within textual records. Future work should aim to extract intensity gradations and a wider variety of climate phenomena. Second, while our operational hypothesis that surviving records represent the most extreme events is pragmatically useful, a more nuanced model incorporating document survival probabilities over time would be beneficial.

Our study establishes a scalable, less labor-intensive pipeline for incorporating historical documents into historical climate reconstruction. Realizing its full potential to extend to other regions and time periods worldwide requires continued international collaboration in excavating, compiling, and encoding new textual archives. Beyond historical climatology, our findings carry even broader implications (Fig. 1B). First, developing multi-modal models that fuse textual records with physical proxies holds great promise for advancing paleoclimatology. Second, our work demonstrates a technical strategy of using physical simulations to enrich sparse, qualitative archives. Third, emerging natural language processing tools, particularly large language models (*47*, *48*), could be applied to further automate and enhance the extraction of climate information from raw historical texts. Moreover, our AI-driven framework for mining historical information serves as a blueprint that can inspire and be adapted by other fields in the humanities and social sciences, including history, economics, and political science, to quantitatively explore the vast, untapped knowledge within our written heritage, bridging the gap between qualitative history and quantitative science.

# Materials and Methods

## 1. Historical documentary data processing

### 1.1. Data source and event definition

Ancient Chinese climate records are dispersed across diverse historical documents, including official histories, local chronicles, imperial memorials, and private notes (*49*). The most comprehensive compilation of these records is *A Compendium of Chinese Meteorological Records in the Last 3,000 Years* (hereinafter *Compendium*) (*50*) published in 2004, a 4-volume work by Zhang De'er and her team based on 7,835 historical sources, covering the period from the 23rd century BC to 1911. We utilize the Reconstructed East Asian Climate Historical Encoded Series (REACHES) dataset constructed by (*16*), which digitalized the *Compendium*.

The core of historical climate records is events, and REACHES addresses this by a comprehensive system of event codes. For example, a record in Vol. 2 of *Compendium* about a drought in Yangzhou, Jiangsu in 1368 says "…On the Dingyou day (a date in the ancient Chinese sexagenary cycle) of the seventh lunar month, Yangzhou Prefecture had no rainfall from the fifth lunar month to this month, with crops damaged by drought (…七月丁酉，扬州府自五月不雨至于是月，旱伤苗稼)". This record corresponds to 3 event codes: 100110189 for no rainfall, 300100009 for drought, and 330112009 for crop failure.

For this study, we defined 4 kinds of climate events: annual flood, annual drought, sub-annual flood, and sub-annual drought, each of which corresponds to a series of REACHES event codes and a specific temporal resolution, respectively (see Table S1 for the complete mapping). Annual flood and drought events are defined as those lacking sub-annual temporal information, and are therefore resolved at the annual level, such as the record under a specific year just saying "drought (旱)" or "severe flood (大涝)". All other events are categorized as sub-annual flood or drought. We name them "sub-annual" rather than "monthly", because the original historical records sometimes only specify seasonal or multi-month timelines rather than precise monthly dates. The aforementioned record of three consecutive rainless months is one such case.

To maintain consistent temporal granularity across the dataset, we disaggregate these seasonally or multi-monthly resolved events into individual months, with one event entry per month in the specified period. Importantly, this monthly formatting reflects a standardized data structure rather than implying that the original records had true monthly precision.

Our event definitions include both meteorological codes (e.g., for no rainfall, persistent rain) and agricultural impact codes (e.g., for crop damage), combining them into a single composite indicator of an extreme climate occurrence. We do not distinguish between meteorological and agricultural flood/drought in our analysis. Our events do not distinguish intensity. Further illustrative examples of historical records, their REACHES codes, and our categorization for both droughts and floods are presented in Tables S2 and S3, respectively.

### 1.2. Calendar convention

Note that ancient Chinese used the Chinese calendar (a lunisolar calendar that combines lunar phases and solar terms) in their records (*51*, *52*), and an important aspect of our methodology involves establishing clear calendar conventions to optimize the alignment between the Chinese and Gregorian calendars (including the proleptic Gregorian calendar on or before 1582). Therefore, we define the following four conventions. See Fig. S1 for their visualization.

First, each lunar month is assumed to have 30 days starting from the Chinese New Year (which varies annually over the period from January 21 to February 20), and a lunar year is defined as 12 lunar months. If a lunar year contains an intercalary month, the last month of that year is excluded.

Second, the Gregorian calendar is similarly standardized with 12 months of 30 days each. Leap days at the end of February and the final 5 days of each Gregorian year are omitted. We refer to our definitions of the Chinese and Gregorian calendars as the "standardized Chinese calendar" and "standardized Gregorian calendar", respectively, and use them throughout this study.

Third, note that ancient Chinese defined seasons by four "solar terms": *Lichun* (立春, around February 4), *Lixia* (立夏, around May 6), *Liqiu* (立秋, around August 8), and *Lidong* (立冬, around November 7) (*52*). In historical documentary records, the four seasons are approximated as 90 days (i.e., 3 months) each following the Chinese New Year in the standardized Chinese calendar.

Finally, to compare our result with the modern climatology, each meteorological season (i.e., MAM, JJA, SON, DJF) is approximated as a 90-day (i.e., 3-month) period starting from the 31st day (i.e., the start of the second month) of the standardized Chinese calendar. See Table S4 for a detailed conversion table for the meteorological seasons.

### 1.3. Spatial gridding of events

A higher spatial resolution demands a greater number of events to ensure robust model training, making it critical to balance spatial resolution against the limited number of events retrievable from historical documents. Following extensive testing to evaluate this trade-off, we adopted a spatial resolution of 1° and 0.57° in the zonal and meridional directions, respectively. This specific resolution was chosen to fully encompass China within a standard 64 × 64 grid, which is a computationally efficient size for the convolutional neural network backbone of our diffusion model. We project each historical event to the nearest grid node. For multiple events assigned to the same grid node and time period, a majority rule resolves conflicts between mutually exclusive types, i.e., annual flood vs. annual drought or sub-annual flood vs. sub-annual drought.

The spatial and temporal distribution of the resulting gridded event dataset is summarized in Figs. S2 and S3, respectively. Spatially, the record density reflects the historical population distribution, with higher counts in the long-settled cores of northern and eastern China and lower counts in the less populated southwestern region. Temporally, the dataset shows a seasonal bias, with the most events recorded in summer and the fewest in winter. The lower event counts in the first ~100 years of the record are consistent with expected documentary loss over time.

## 2. Climate model data for training and validation

### 2.1. Data source and dataset split

To meet the substantial data requirements of training diffusion models, which far exceed the volume of available observational records, we turned to climate model simulations. We therefore incorporate simulated data from the Community Earth System Model version 1 with the Community Atmospheric Model version 5 (CESM1-CAM5) (*17*, *18*). This dataset encompasses two scenarios: RCP4.5 and RCP8.5, which are two potential trajectories of global warming with projected radiative forcing levels of 4.5 W/m² and 8.5 W/m² by 2100, respectively.

Each ensemble member begins from a slightly perturbed initial state on January 1, 1920, and integrates forward under time-evolving external forcings (e.g., greenhouse gases, aerosols, solar irradiance, and volcanic aerosols). The integration follows historical forcings until 2005 and then diverges into the RCP4.5 and RCP8.5 scenario pathways until 2080 and 2100, respectively. Critically, our primary objective in this study is to train a model to reconstruct the spatial patterns of precipitation anomalies associated with documented extreme events, a task that relies more on learning the model's internal climate variability than its forced response to specific historical forcings. Therefore, this combination of historical and scenario simulations serves as an efficient method to generate a large, diverse sample of the climate model's internal variability under varying background states, which is essential for robust AI model training. From the available ensembles, we selected 5 ensemble members from each scenario for training, and held out 1 ensemble member from RCP4.5 for validation, corresponding to 1,710 years of training data and 161 years of validation data. To further test the robustness of our reconstruction to the choice of training data, we conducted an ablation study using the data from the pre-industrial control simulation of Institut Pierre-Simon Laplace Coupled Model, version 6A, Low Resolution (IPSL-CM6A-LR) (see Supplementary Text Section 9.1).

## 2.2. Data augmentation and normalization

Note that for data augmentation, we allow annual time series to start on any calendar day (e.g., January 1, June 8, etc.), with the definition of years and months adhering to the standardized Gregorian calendar as established earlier. All the years starting at the same month and day form an independent year group, and the same normalization procedure is applied within each group. Consequently, the "years" and "months" used for training do not necessarily correspond to calendar years or months, but still adhere to the 30-day monthly and 360-day annual structure of the standardized Gregorian calendar. We call them "custom years" and "custom months" hereinafter.

To remove long-term climate change trends and align with the historical recorders' perception where the significance of an event is determined by comparison with other contemporaneous similar events, all precipitation data were normalized using statistics from a 31-year sliding window, including the target year plus 15 years preceding and 15 years following it. For the years at the start or end of the data sequence where a full 31-year window is unavailable, the window is truncated to the beginning or end of the sequence, respectively. This sliding-window normalized precipitation is referred to as "normalized precipitation" for brevity in the subsequent sections. The choice of window size was inspired by climatological conventions, as 30 years is typically adopted as a standard climatological reference period.

## 2.3. Calibration of events

We operate under the hypothesis that historical documents primarily preserve the most extreme climate events. Therefore, we calibrate our training and validation datasets such that the number of events at each grid point is consistent with those recorded in the historical documents. The algorithm is detailed as follows.

First, we quantified the occurrence frequency of each event category in the historical documents as defined earlier. For annual flood and drought events, we calculated the average number of such events per year for each grid point. For sub-annual flood and drought events, we calculated an annual average for events in each month for each grid point, such that each month and grid point has a corresponding annual average number of events. Note that as the annual time series in the training dataset is flexible to start on any calendar date, the event frequency varies with different start dates and the transition should be smooth. For this end, we process the historically recorded events as follows: each event is projected to the days of its calendar year or month, and then we derive the event count for each custom year and month by summing the events whose projected days fall within the corresponding temporal boundaries.

Second, within the training and validation datasets, we rank the normalized precipitation values separately for each grid point for the annual and monthly time series. For annual normalized precipitation, which corresponds to annual flood and drought events, we select a subset of values matching the grid-point-specific number of the annual flood and drought events recorded in historical documents. Specifically, we choose the top-ranked values for flood events and the bottom-ranked values for drought events at each grid point, with this selection applied to each ensemble member for the training and validation datasets. An analogous approach is applied to sub-annual flood and drought events: we rank the monthly normalized precipitation values for each grid point, and the number of selected values corresponds to the grid-point-specific number of sub-annual events recorded in historical documents for each month. This grid-point-wise calibration ensures that the synthesized training events not only match the global historical event totals but also replicate their observed spatial distribution.

Below is a brief pseudocode describing the processing for training and validation datasets.

**Algorithm 1**. Pseudocode for processing the training and validation datasets.

1: **for** each ensemble member **in** the training and validation datasets **do**
  2: **for** each unique start date (e.g., January 1, June 8) **do**
    3: Define "custom year" and "custom month" following the standardized Gregorian calendar (30 days/month, 360 days/year), such that the ensemble member is split into multiple custom years/months based on the start date
    4: Project all historical events to specific calendar days and grid points ($i,j$)
    5: Derive event counts for each custom year, month, and grid point ($i,j$) by summing events associated with their corresponding days and locations
    # The following statistics are annual average event counts calculated for each grid point ($i,j$) from historical records
    6: Calculate $Naf(i,j)$ = Annual average number of custom-year annual flood events at grid point ($i,j$)
    7: Calculate $Nad(i,j)$ = Annual average number of custom-year annual drought events at grid point ($i,j$)
    8: Calculate $Nsf(i,j,m)$ = Monthly average number of custom-month sub-annual flood events for month $m$ at grid point ($i,j$)
    9: Calculate $Nsd(i,j,m)$ = Monthly average number of custom-month sub-annual drought events for month $m$ at grid point ($i,j$)
    # The following part processes normalized annual and monthly precipitation using statistics from the 31-year sliding window (truncated if a full 31-year window is unavailable)
    # For processing normalized annual precipitation
    10: **for** each grid point ($i,j$) **do**
      11: Select the top $Naf(i,j)$ and bottom $Nad(i,j)$ ranked values as annual flood/drought events for grid point ($i,j$)

```
    12: end for
    # For processing normalized monthly precipitation
    13: for each grid point (i,j) do
       14: for each month m in range(12) do
          15: Select the top Nsf(i,j,m) and bottom Nsd(i,j,m) ranked values as sub-annual
              flood/drought events for month m at grid point (i,j)
       16: end for
    17: end for
  18: end for
19: end for
```

## 3. Diffusion model: Architecture, training, and inference

### 3.1. Model architecture

As outlined in the main text, we employ a 3D diffusion model, an architecture renowned in video generation for its robustness to noise and overfitting (*11*, *23*). Its inherent denoising process, which begins from random Gaussian noise, not only facilitates massive sampling but also is naturally suited for ensemble-based uncertainty quantification. We implement the model using the *UNet3DConditionModel* from the Python *diffusers* package (*53*). The model backbone is a U-Net with a symmetric encoder-decoder structure, comprising 4 down-sampling and 4 up-sampling blocks, each containing 4 layers. To effectively handle spatiotemporal data, the architecture processes the temporal dimension distinctly from the spatial dimensions. As illustrated in Fig. S7, a temporal self-attention mechanism is incorporated before the encoder to capture long-range dependencies across time. Furthermore, each layer within the encoder and decoder is equipped with a temporal convolution layer to model short-range temporal correlations.

A critical component of our design is conditioning the model on the start date of the input series. This serves two purposes: First, it acts as a data augmentation technique to allow the model to be trained on annual series starting from any calendar day, which is essential for the substantial data requirements of diffusion models. Second, it enables the model to account for the variable timing of the Chinese New Year in the Gregorian calendar during inference. Given a start date $(y, m, d)$, we construct a periodic conditioning prompt as a 2D vector

$$(\sin 2\pi d_s, \cos 2\pi d_s) = \left(\sin 2\pi \frac{\text{DoY}(y,m,d)}{\text{DiY}(y)}, \cos 2\pi \frac{\text{DoY}(y,m,d)}{\text{DiY}(y)}\right),$$

where $\text{DoY}(\cdot)$ and $\text{DiY}(\cdot)$ represent the day of the year and the number of days in the year, respectively. To guide the model in learning large-granularity climate patterns, we inject the prompt via cross-attention into the bottleneck of the U-Net with the highest-level feature representations, namely the final encoder block, the middle block, and the first decoder block. The model is trained directly in the raw data space. Consistent with findings from previous studies, we observed that the *v*-prediction parameterization yields better numerical convergence than the classical *ε*-prediction in our application (*54*, *55*).

### 3.2. Input and output representation

The model input is a 4D tensor of shape $(N, C_{\text{in}}, H, W) = (12, 3, 64, 64)$, representing 12 months, 3 input channels, and a 64×64 spatial grid. The three input channels consist of a noise channel and two event channels. The first event channel encodes annual flood and drought events, while the second encodes sub-annual events. Both event channels use a ternary scheme: 1 for flood, −1 for drought, and 0 for no event. Since annual events are constant for all months within a year, the values in the annual event channel are replicated across the temporal dimension. The output is a tensor of shape ($N$, $C_{out}$, $H$, $W$) = (12, 1, 64, 64), corresponding to the normalized precipitation for each month.

### 3.3. Training configuration

Training was conducted using a mixed-precision of fp16 on 40 Huawei Ascend 910B AI accelerators. We used the Adam optimizer ($\beta_1 = 0.9$, $\beta_2 = 0.999$) with a learning rate of $5 \times 10^{-4}$, a linear warm-up of 1000 steps, a batch size of 320, and 1000 diffusion steps. The model was trained for 20 epochs on a dataset of 515k samples, with each epoch requiring approximately 50 minutes. Convergence was achieved after 20 epochs as evidenced by the plateauing of validation metrics (see Table S5).

### 3.4. Inference and post-processing

During inference, we input the historical documentary data and convey the timing of the Chinese New Year to the model via the periodic conditioning prompt. Correspondingly, the model outputs the normalized precipitation for the target lunar year. 80 samples were generated per year as the ensemble output, with their median selected as the deterministic result for subsequent analysis. For reference, with 200 diffusion steps, each Huawei Ascend 910B AI accelerator card can infer a batch of 30 samples in approximately 2 minutes.

The raw model outputs, which are the normalized monthly precipitation, were first denormalized using the mean and standard deviation of precipitation from 1920 to 1935. We choose this period as it is the earliest available in our model data and thus represents a pre-industrial climate state that is closest, albeit not perfect, to the Ming–Qing period. These denormalized monthly values were then aggregated into seasonal and annual totals by summing across the corresponding months. Next, the mean and standard deviation of the deterministic seasonal and annual precipitation were computed over the entire Ming–Qing period, and these statistics were used to re-normalize both the deterministic and ensemble results, ensuring consistency with the historical climate context. All subsequent result demonstrations and analyses are based on these re-normalized precipitation data. The robustness of our conclusions to this choice of re-normalization is confirmed through an ablation study using an alternative rank-based sliding window algorithm (see Supplementary Text Section 9.2).

## 4. Validation statistics

Our diffusion model intrinsically provides an ensemble output for uncertainty quantification (UQ). For practical application and analysis, we define the ensemble median as our deterministic reference. We therefore validate our model from both deterministic and ensemble perspectives using four statistics: Square of the Pearson correlation coefficient (RSQ) and coefficient of efficiency (CE) for the deterministic evaluation, and continuous ranked probability skill score (CRPSS) and spread-skill ratio (SSR) for the ensemble evaluation. The definitions and interpretations of these metrics are provided below.

RSQ is commonly used in climatology and meteorology (*24*, *25*). It is calculated as

$$\text{RSQ}_{i,j} = \frac{\left[\sum_t (X_{t,i,j} - \bar{X}^v_{i,j})(\hat{X}_{t,i,j} - \bar{\hat{X}}^v_{i,j})\right]^2}{\sum_t (X_{t,i,j} - \bar{X}^v_{i,j})^2 \sum_t (\hat{X}_{t,i,j} - \bar{\hat{X}}^v_{i,j})^2},$$

where $X_{t,i,j}$ and $\hat{X}_{t,i,j}$ are the ground truth and deterministic reconstructed values at time $t$ and spatial coordinate $(i,j)$, respectively, and $\bar{X}^v_{i,j}$ and $\bar{\hat{X}}^v_{i,j}$ are the temporal mean of $X_{t,i,j}$ and $\hat{X}_{t,i,j}$ at the validation dataset, respectively. Its value range is [0, 1], where 1 means perfect correlation and 0 means no correlation.

CE indicates whether the deterministic reconstruction performs better than the climatology on the validation set (*24*, *25*). It is defined as

$$\text{CE}_{i,j} = 1 - \left[\frac{\sum_t (X_{t,i,j} - \hat{X}_{t,i,j})^2}{\sum_t (X_{t,i,j} - \bar{X}^v_{i,j})^2}\right].$$

Its value range is $(-\infty, 1]$, where a non-positive value means the reconstruction performs no better than the climatology, thus a positive value is necessary to demonstrate model skill. Note that reduction of error (RE) is defined similarly but the climatology is calculated on the training set (*24*, *25*). As the training and validation data are processed similarly and share the same statistics in our study, RE = CE.

The first two metrics evaluate the deterministic reconstruction, while the following two assess the ensemble output. Continuous ranked probability score (CRPS) is widely used in ensemble weather forecasting (*56*, *57*). It is defined as

$$\text{CRPS}_{t,i,j} = \int_{-\infty}^{\infty} [F(\widehat{\mathbf{X}}_{t,i,j}) - \mathbf{1}(X_{t,i,j} \leq z)]^2 dz,$$

where $\widehat{\mathbf{X}}_{t,i,j}$ is the ensemble reconstructed value (thus $\hat{X}_{t,i,j}$ is the median of $\widehat{\mathbf{X}}_{t,i,j}$), and $F(\cdot)$ and $\mathbf{1}(\cdot)$ are the cumulative distribution and Heaviside step functions, respectively. Based on CRPS, we define CRPSS as

$$\text{CRPSS} = \frac{\widetilde{\text{CRPS}} - \widehat{\text{CRPS}}}{\widetilde{\text{CRPS}}},$$

where $\widehat{\text{CRPS}}$ and $\widetilde{\text{CRPS}}$ are the CRPS for the reconstruction and climatology, respectively. CRPSS has the same value range with CE of $(-\infty, 1]$, where a non-positive value also means the reconstruction is worse than the climatology. As we have normalized our precipitation to the standard normal distribution, $\widetilde{\text{CRPS}}$ has a theoretical value of $\frac{1}{\sqrt{\pi}} = 0.5642$.

SSR is defined as

$$\text{SSR}_t = \frac{\left[\text{mean}_{i,j}\left(\text{var}(\widehat{\mathbf{X}}_{t,i,j})\right)\right]^{1/2}}{\left[\text{mean}_{i,j}\left(\text{mean}_e(\widehat{\mathbf{X}}_{t,i,j}) - X_{t,i,j}\right)^2\right]^{1/2}},$$

where $\text{mean}_{i,j}(\cdot)$ and $\text{mean}_e(\cdot)$ mean spatial and ensemble average, respectively (*7*). As its name suggests, it is the ratio of spread and skill, where spread is the standard deviation of the ensemble reconstruction and skill is the root-mean-square error between the ensemble reconstruction and ground truth. A perfect reconstruction has an SSR of 1, and a value smaller and larger than 1 mean the reconstruction is over- and under-confident, respectively.

## 5. Consistency alignment of multi-source ENSO 3.4 indices

To compare the ENSO-precipitation correlations between our historical reconstruction (1368–1911) and the modern period (1940–2024, based on ERA5 reanalysis), a consistent ENSO 3.4 index across both eras was required. We employed the prior winter (November–January, NDJ) index from (*45*), which spans 1301–2005 and is referenced to the 1971–2000 average. This index was extended to 2024 using the National Oceanic and Atmospheric Administration's (NOAA) Oceanic Niño Index (ONI) (*58*). We extracted the NDJ values from the NOAA ONI's 3-month running means to match the temporal definition of (*45*), normalized them to the same 1971–2000 baseline, and applied a linear regression to calibrate the two series over their overlapping period (1950–2005), thereby ensuring a continuous and consistent ENSO 3.4 record.

# Supplementary Text

This section presents supplementary results to validate our reconstruction and demonstrate its potential applications. It also includes ablation studies that test the robustness of our key methodological choices.

## 6. Model validation and skill assessment

We first report the quantitative skill of our model on the held-out validation dataset. The values of the validation statistics for our model, namely RSQ, CE, CRPSS, and SSR, are summarized in Table S6. For our inferred annual precipitation on the 161-year validation dataset, the RSQ and CE are 0.61 and 0.70, respectively, which are comparable to those reported by (*25*). In terms of ensemble statistics, our forecasts show a CRPSS of 0.45, and the SSR is 1.01. All statistics indicate strong model performance for annual precipitation. For seasonal precipitation, the average RSQ is 0.42, with summer exhibiting the highest value (0.48) and winter the lowest (0.38). At the monthly scale, the average RSQ decreases to 0.35, with May having the highest value (0.41) and January the lowest (0.29). The complete dashboard and spatial distributions of these statistics are provided in Table S6 and Figs. S4–S6. The deterioration in reconstruction quality with increasing temporal resolution is expected, as it is inherently linked to sparser historical records and greater precipitation variability at finer timescales. Despite these challenges, our model demonstrates remarkable skill in reconstructing seasonal precipitation and moderate skill in reconstructing monthly precipitation. The model skill is lower in winter than in summer for similar reasons (Fig. S2). These limitations highlight the intrinsic challenge of reconstructing high-frequency climate from sparse archives and point to the need for future efforts to uncover additional historical records, particularly for winter, to further constrain the model and reduce uncertainties.

## 7. Cross-validation against independent reconstructions

We further cross-validate our historical precipitation reconstruction derived from historical climate records against the reconstruction of (*25*), which uses a classical PPR-based algorithm and integrates both historical documents and tree-ring proxies. A key methodological distinction is that we assume surviving historical events represent the most extreme conditions within a local time window, whereas (*25*) categorized flood and drought events by intensity to derive a dryness/wetness index (DWI). To align with their analysis, we focus on precipitation from May to September (MJJAS), the core rainy season in eastern China. As shown in Fig. S8, despite this methodological divergence, the correlation between the two reconstructions exceeds 0.5 across most of the study area and reaches up to 0.75 in data-rich northern China. Conversely, lower correlations (< 0.5) occur in the southwestern region, where both Figs. S2 and 1A of (*25*) and our dataset (Fig. S2) lack sufficient data coverage. We then perform an Empirical Orthogonal Function (EOF) analysis on the same MJJAS precipitation. As shown in Fig. S9, our leading three modes are a uniform mode, a meridional dipolar mode, and a meridional tripolar mode, respectively, agreeing with Fig. 8 of (*25*) on both the general spatial patterns and the location of the boundaries between the adjacent poles.

## 8. Supplementary analyses of the reconstruction

Supplementary results for the EOF analysis of our reconstruction are presented herein. Figure S10 shows the top three years with extreme high and low principal component (PC) values corresponding to the first four EOF modes of the annual precipitation. Specifically, 1593 and 1877 are among the years with extreme high and low PC values for the uniform mode, respectively. Figure S11 presents the EOF analysis of the seasonal precipitation. Across all the seasons, the two leading EOF modes are a uniform spatial mode and a meridional dipolar mode. In summer, autumn and winter, the uniform mode precedes the meridional dipolar mode, consistent with the EOF modes of annual precipitation, whereas the meridional dipolar mode precedes the uniform mode in spring. The subsequent two modes for all the seasons are characterized as a meridional tripolar mode and a zonal bipolar mode, analogous to those derived from the annual precipitation. Notably, the explained variances of the uniform mode in winter are significantly higher than those in the other seasons.

Figure S12 presents the averaged normalized annual and seasonal precipitation in China's five major river basins in Ming and Qing Dynasties (1368–1911). From north to south, these basins are the Hai, Yellow, Huai, Yangtze, and Pearl Rivers, where Hai, Yellow and Huai River Basins are generally considered as a part of northern China, and Yangtze and Pearl River Basins belong to southern China. Note that the basin-averaged values are calculated by first normalizing precipitation at each grid point and then averaging these normalized values within each river basin boundary. This differs from the alternative approach of averaging raw precipitation in the basin first followed by normalization. This method assigns equal weight to each grid point instead of weighting by grid-specific mean precipitation, emphasizing the spatial coverage of climate events. The top three flood and drought years for each basin are highlighted in the figure, and their corresponding basin-averaged normalized annual precipitation values are further detailed in Table S7. The Chongzhen Drought (1637–1643) and Guangxu Drought (1875–1879) are two of the most significant historical drought events in Chinese history and have been extensively studied in historical climatology (*27–31*). Their peak years 1640 and 1877 correspond to the most severe droughts on record in the study period. Specifically, the year 1640 ranks second in drought intensity for the Huai River Basin, and the year 1877 ranks first and second in drought intensity for the Yellow and Hai River Basin, respectively. Our model provides ensemble-based uncertainty estimates, quantified as the standard deviation of the basin average (Table S7). This uncertainty is influenced by the density of documented events and the basin area. Basins in regions with fewer records (e.g., the Pearl River Basin) show higher uncertainty, whereas larger basins (e.g., the Yangtze River Basin) exhibit lower uncertainty because their averages integrate over more grid points, reducing the standard error of the mean. By providing such spatially resolved quantification, our reconstruction offers a robust tool for assessing how extreme climate anomalies differentially impacted historical developments across regions. It should be noted that the global re-normalization method, as introduced in Materials and Methods Section 3.4, results in dampened anomaly amplitudes in the early part of the record due to lower documentary coverage. The implications of this choice and an alternative re-normalization method designed to mitigate it are evaluated in the subsequent ablation study.

To enhance the understanding of the development of the Chongzhen and Guangxu Droughts, including their initiation, peak intensity, and eventual cessation, we further analyzed their spatial distributions and statistical characteristics. Figure S13 presents the spatial distribution of normalized annual and seasonal precipitation during the Chongzhen Drought. Drought intensity in northern China was greater before 1640 than after 1640. While the drought was predominantly concentrated in northern China, the drought in southern China in 1643 is also prominent. Figure S14 quantifies the regional impacts of the Chongzhen Drought using basin-averaged normalized seasonal precipitation in China's five major river basins. The most severe phase of the drought in northern China occurred from the spring to the autumn of 1640, with normalized seasonal precipitation values of approximately −2 in the Hai, Yellow and Huai River Basins. For southern China, the most severe drought conditions emerged in the summer and autumn of 1643, with normalized seasonal precipitation around −1.5 in the Yangtze and Pearl River Basins.

Figure S15 presents the spatial distribution of normalized annual and seasonal precipitation during the Guangxu Drought. This drought initiated in northern China in the spring of 1875, with its most severe phase occurring from the spring of 1877 to the spring of 1878. Drought conditions began to alleviate from the summer of 1878 and concluded in 1879. These developmental processes are also reflected in the variations of normalized seasonal precipitation during the Guangxu Drought as shown in Fig. S16.

While the uncertainty in our monthly precipitation reconstruction is higher than at coarser timescales, the data nevertheless offer valuable insights into sub-seasonal climate dynamics. To illustrate this utility, we present an example using the reconstructed monthly precipitation patterns for the year 1721 (Fig. S17). As shown in the figure, the annual precipitation pattern indicates a prominent widespread drought across northern, central, and southern China, though no famine was triggered due to the robust national strength during the Kangxi reign. For seasonal and monthly precipitation, the shift in the region of above-average precipitation is notable. This region shifted sequentially along the Yangtze River, Southwestern China, the coastal areas of southern China, and northern China. It finally extended nationwide in the late year, with precipitation particularly above average over the coastal areas of Southwestern China and along the Huai River. Therefore, while these monthly reconstructions are suitable for identifying patterns and conducting semi-quantitative analyses, users should be mindful of their elevated uncertainty when interpreting precise quantitative values.

## 9. Ablation studies and robustness tests

To evaluate the robustness of our primary findings, we conducted a suite of three ablation studies to test the alternatives of key assumptions and data imperfections. These tests move from choices in data processing to the fundamental challenge of sparse historical records.

### 9.1. Sensitivity to climate model training data

We replaced the primary training data, which consists of CESM1-CAM5 under future forcing scenarios, with a long unforced control simulation from the IPSL-CM6A-LR pre-industrial control (piControl) experiment (*59*). This piControl simulation is a 2000-year run with constant pre-1850 forcing conditions, designed to model internal climate variability in the absence of anthropogenic change. To maintain a consistent 20th-century climate baseline with our primary training data and avoid temporal overlap with our reconstruction period, we discarded the first 70 model years (1850–1919) and used the subsequent 1,930-year segment (model years 1920–3849) for training. The resulting reconstruction (Fig. S18) preserves the overall temporal evolution and anomaly amplitudes of the primary basin-averaged series, with substantial overlap in the identified most extreme years. The recovered leading climate modes (Fig. S19) and the multi-century ENSO teleconnection pattern (Fig. S20) are also consistent. This demonstrates that our primary reconstruction is not an artifact of a single model's climatology or its forced response. A detailed comparison of Figs. S19A and S19B shows the generative model synthesizes the information from the training data, as certain spatial features in the reconstruction reflect characteristics present in its specific training data.

### 9.2. Sensitivity to re-normalization method

To address the dampening of early-record anomalies identified above, we tested an alternative precipitation re-normalization algorithm. We applied a rank-based sliding window re-normalization instead of the primary re-normalization algorithm described in Materials and Methods Section 3.4. In this method, precipitation within each 31-year sliding window is ranked and mapped to a z-score via its empirical cumulative distribution function, and the final re-normalized value for a given year is the average of its z-scores across all windows containing it. Compared to the primary re-normalization, this method mitigates the dampening of early-record anomalies intrinsic to sparse data but reduces the relative intensity of the most extreme events across the full time series (Fig. S21), illustrating a trade-off in re-normalization strategy. Despite this difference in scaling, the recovered large-scale climate modes (Fig. S22) and the ENSO teleconnection patterns (Fig. S23) remain consistent. This confirms that our core climatic conclusions are robust to the choice of re-normalization strategy for handling uneven data density over time.

### 9.3. Sensitivity to random gaps in the historical archive

To assess the robustness of our reconstruction to gaps in the historical record, we performed an ablation study that randomly discards a fraction of documented events. We tested two dropout ratios, 0.2 and 0.5. The basin-averaged normalized annual and seasonal precipitation under these scenarios (Figs. S24 and S25) show that both the amplitude of climate fluctuations and the identity of the most extreme flood and drought years are largely preserved. The leading EOF patterns and ENSO-precipitation teleconnections (not shown) remain virtually unchanged, confirming that our core results are robust to significant random data loss.

**References**

1. U. Büntgen, W. Tegel, K. Nicolussi, M. McCormick, D. Frank, V. Trouet, J. O. Kaplan, F. Herzig, K.-U. Heussner, H. Wanner, J. Luterbacher, J. Esper, 2500 years of European climate variability and human susceptibility. *Science* **331**, 578–582 (2011).

2. E. R. Cook, R. Seager, R. R. Heim, R. S. Vose, C. Herweijer, C. Woodhouse, Megadroughts in North America: Placing IPCC projections of hydroclimatic change in a long-term context. *Journal of Quaternary Science* **25**, 48–61 (2010).

3. N. Pederson, A. E. Hessl, N. Baatarbileg, K. J. Anchukaitis, N. Di Cosmo, Pluvials, droughts, the Mongol Empire, and modern Mongolia. *Proceedings of the National Academy of Sciences* **111**, 4375–4379 (2014).

4. P. Braconnot, S. P. Harrison, M. Kageyama, P. J. Bartlein, V. Masson-Delmotte, A. Abe-Ouchi, B. Otto-Bliesner, Y. Zhao, Evaluation of climate models using palaeoclimatic data. *Nature Climate Change* **2**, 417–424 (2012).

5. G. A. Schmidt, J. D. Annan, P. J. Bartlein, B. I. Cook, E. Guilyardi, J. C. Hargreaves, S. P. Harrison, M. Kageyama, A. N. LeGrande, B. Konecky, S. Lovejoy, M. E. Mann, V. Masson-Delmotte, C. Risi, D. Thompson, A. Timmermann, L.-B. Tremblay, P. Yiou, Using palaeo-climate comparisons to constrain future projections in CMIP5. *Climate of the Past* **10**, 221–250 (2014).

6. PAGES 2k Consortium, Continental-scale temperature variability during the past two millennia. *Nature Geoscience* **6**, 339–346 (2013).

7. K. Bi, L. Xie, H. Zhang, X. Chen, X. Gu, Q. Tian, Accurate medium-range global weather forecasting with 3D neural networks. *Nature* **619**, 533–538 (2023).

8. J. Pathak, S. Subramanian, P. Harrington, S. Raja, A. Chattopadhyay, M. Mardani, T. Kurth, D. Hall, Z. Li, K. Azizzadenesheli, P. Hassanzadeh, K. Kashinath, A. Anandkumar, FourCastNet: A global data-driven high-resolution weather model using adaptive Fourier neural operators. arXiv:2202.11214 [physics.ao-ph] (2022).

9. R. Lam, A. Sanchez-Gonzalez, M. Willson, P. Wirnsberger, M. Fortunato, F. Alet, S. Ravuri, T. Ewalds, Z. Eaton-Rosen, W. Hu, A. Merose, S. Hoyer, G. Holland, O. Vinyals, J. Stott, A. Pritzel, S. Mohamed, P. Battaglia, GraphCast: Learning skillful medium-range global weather forecasting. *Science* **382**, 1416–1421 (2023).

10. R. Rombach, A. Blattmann, D. Lorenz, P. Esser, B. Ommer, "High-resolution image synthesis with latent diffusion models" in *Proceedings of the IEEE/CVF Conference on Computer Vision and Pattern Recognition (CVPR)* (2022), pp. 10684–10695.

11. J. Ho, A. N. Jain, P. Abbeel, Denoising diffusion probabilistic models. *Advances in Neural Information Processing Systems* **33**, 6840–6851 (2020).

12. J. Jumper, R. Evans, A. Pritzel, T. Green, M. Figurnov, O. Ronneberger, K. Tunyasuvunakool, R. Bates, A. Žídek, A. Potapenko, A. Bridgland, C. Meyer, S. A. A. Kohl, A. J. Ballard, A. Cowie, B. Romera-Paredes, S. Nikolov, R. Jain, J. Adler, T. Back, S. Petersen, D. Reiman, E. Clancy, M. Zielinski, M. Steinegger, M. Pacholska, T. Berghammer, S. Bodenstein, D. Silver, O. Vinyals, A. W. Senior, K. Kavukcuoglu, P. Kohli, D. Hassabis, Highly accurate protein structure prediction with AlphaFold. *Nature* **596**, 583–589 (2021).

13. W. Qi, S. Liu, M. Zhao, Z. Liu, China's different spatial patterns of population growth based on the "Hu Line". *Journal of Geographical Sciences* **26**, 1611–1625 (2016).

14. National Bureau of Statistics of China, *Major Figures on 2020 Population Census of China* (National Bureau of Statistics of China, Beijing, 2021).

15. Resource and Environment Science Data Platform, *Nine major river basins of China*, Resource and Environment Science Data Platform (2025); https://www.resdc.cn/data.aspx?DATAID=141.

16. P. K. Wang, K. H. E. Lin, Y.-C. Liao, H.-M. Liao, Y.-S. Lin, C.-T. Hsu, S.-M. Hsu, C.-W. Wan, S.-Y. Lee, I-C. Fan, P.-H. Tan, T.-T. Ting, Construction of the REACHES climate database based on historical documents of China. *Scientific Data* 5, 180288 (2018).

17. B. M. Sanderson, K. W. Oleson, W. G. Strand, F. Lehner, B. C. O'Neill, A new ensemble of GCM simulations to assess avoided impacts in a climate mitigation scenario. *Climatic Change* **146**, 303–318 (2018).

18. J. W. Hurrell, M. M. Holland, P. R. Gent, S. Ghan, J. E. Kay, P. J. Kushner, J.-F. Lamarque, W. G. Large, D. Lawrence, K. Lindsay, W. H. Lipscomb, M. C. Long, N. Mahowald, D. R. Marsh, R. B. Neale, P. Rasch, S. Vavrus, M. Vertenstein, D. Bader, W. D. Collins, J. J. Hack, J. Kiehl, S. Marshall, The Community Earth System Model: A Framework for Collaborative Research. *Bulletin of the American Meteorological Society* **94**, 1339–1360 (2013).

19. Z. Wang, Y. Li, B. Liu, J. Liu, Global climate internal variability in a 2000-year control simulation with Community Earth System Model (CESM). *Chinese Geographical Science* **25**, 263–273 (2015).

20. Q. Yan, Z. Zhang, H. Wang, D. Jiang, Simulated warm periods of climate over China during the last two millennia: The Sui-Tang warm period versus the Song-Yuan warm period. *Journal of Geophysical Research: Atmospheres* **120**, 2229–2241 (2015).

21. Z. Jiang, W. Li, J. Xu, L. Li, Extreme precipitation indices over China in CMIP5 models. Part I: Model evaluation. *Journal of Climate* **28**, 8603–8619 (2015).

22. WMO Working Group, "WMO guidelines on the calculation of climate normals" (WMO-No. 1203, World Meteorological Organization, Geneva, 2017).

23. J. Leinonen, U. Hamann, D. Nerini, U. Germann, G. Franch, Latent diffusion models for generative precipitation nowcasting with accurate uncertainty quantification. arXiv:2304.12891 [physics.ao-ph] (2023).

24. E. R. Cook, K. J. Anchukaitis, B. M. Buckley, R. D. D'Arrigo, G. C. Jacoby, W. E. Wright, Asian monsoon failure and megadrought during the last millennium. *Science* **328**, 486–489 (2010).

25. F. Shi, S. Zhao, Z. Guo, H. Goosse, Q. Yin, Multi-proxy reconstructions of May–September precipitation field in China over the past 500 years. *Climate of the Past* **13**, 1919–1938 (2017).

26. Y. Wu, H. Yao, G. Wang, G. Shen, R. Shi, B. Hou, Analysis on characteristics of extreme drought and flood events in Huaihe River Basin. *Hydro-Science and Engineering* **4**, 149–153 (2011).

27. J. B. Parsons, *The Peasant Rebellions of the Late Ming Dynasty* (University of Arizona Press, Tucson, AZ, 1970).

28. C. Shen, W.-C. Wang, Z. Hao, W. Gong, Exceptional drought events over eastern China during the last five centuries. *Climatic Change* **85**, 453–471 (2007).

29. M. Davis, *Late Victorian holocausts: El Niño famines and the making of the third world* (Verso, London and New York, 2001).

30. X. Zhai, X. Fang, Y. Su, Regional interactions in social responses to extreme climate events: A case study of the North China famine of 1876–1879. *Atmosphere* **11**, 393 (2020).

31. K. Edgerton-Tarpley, Family and gender in famine: Cultural responses to disaster in North China, 1876–1879. *Journal of Women's History* **16**, 119–147 (2004).

32. Z. Hao, J. Zheng, G. Wu, X. Zhang, Q. Ge, 1876–1878 severe drought in North China: Facts, impacts and climatic background. *Chinese Science Bulletin* **55**, 3001–3007 (2010).

33. Y. Ding, J. C. L. Chan, The East Asian summer monsoon: an overview. *Meteorology and Atmospheric Physics* **89**, 117–142 (2005).

34. T. Sampe, S.-P. Xie, Large-scale dynamics of the meiyu-baiu rainband: Environmental forcing by the westerly jet. *Journal of Climate* **23**, 113–134 (2010).

35. Y. Ding, Z. Wang, Y. Sun, Inter-decadal variation of the summer precipitation in East China and its association with decreasing Asian summer monsoon. Part I: Observed evidences. *International Journal of Climatology* **28**, 1139–1161 (2008).

36. Y. Duan, Q. Yang, Z. Ma, P. Wu, X. Chen, J. Duan, Disentangling the driving mechanisms of the tripole mode of summer rainfall over eastern China. *Journal of Climate* **36**, 1175–1186 (2023).

37. C. He, A. Lin, D. Gu, C. Li, B. Zheng, T. Zhou, Interannual variability of eastern China summer rainfall: The origins of the meridional triple and dipole modes. *Climate Dynamics* **48**, 683–696 (2017).

38. B. Wu, X. Lang, D. Jiang, Changes in Summer Precipitation Modes over Eastern China in Simulated Warm Intervals of the Last Interglacial, Mid-Holocene, and Twenty-First Century. *Journal of Climate* **36**, 2401–2420 (2023).

39. F. Shi, H. Goosse, F. Klein, S. Zhao, T. Liu, Z. Guo, Monopole mode of precipitation in East Asia modulated by the South China Sea over the last four centuries. *Geophysical Research Letters* **46**, 14713–14722 (2019).

40. W. H. Qian, Q. Hu, Y. F. Zhu, D. K. Lee, Centennial-scale dry-wet variations in East Asia. *Climate Dynamics* **21**, 77–89 (2003).

41. H. Fu, F. Shi, W. Liu, H. Xue, W. Man, J. Li, Z. Guo, Tracing the centennial variation of East Asian Summer Monsoon. *Global and Planetary Change* **238**, 104464 (2024).

42. H. Shi, B. Wang, E. R. Cook, J. Liu, F. Liu, Asian summer precipitation over the past 544 years reconstructed by merging tree rings and historical documentary records. *Journal of Climate* **31**, 7845–7861 (2018).

43. T. Lan, X. Yan, Analysis of drought characteristics and causes in Yunnan province in the last 60 years (1961–2020). *Journal of Hydrometeorology* **25**, 177–190 (2024).

44. Y. Liu, D. Yan, A. Wen, Z. Shi, T. Chen, R. Chen, Relationship between precipitation characteristics at different scales and drought/flood during the past 40 years in Longchuan river, Southwestern China. *Agriculture* **12**, 89 (2022).

45. J. Li, S.-P. Xie, E. R. Cook, M. S. Morales, D. A. Christie, N. C. Johnson, F. Chen, R. D'Arrigo, A. M. Fowler, X. Gou, K. Fang, El Niño modulations over the past seven centuries. *Nature Climate Change* **3**, 822–826 (2013).

46. H. Hersbach, B. Bell, P. Berrisford, S. Hirahara, A. Horányi, J. Muñoz-Sabater, J. Nicolas, C. Peubey, R. Radu, D. Schepers, A. Simmons, C. Soci, S. Abdalla, X. Abellan, G. Balsamo, P. Bechtold, G. Biavati, J. Bidlot, M. Bonavita, G. De Chiara, P. Dahlgren, D. Dee, M. Diamantakis, R. Dragani, J. Flemming, R. Forbes, M. Fuentes, A. Geer, L. Haimberger, S. Healy, R. J. Hogan, E. Hólm, M. Janisková, S. Keeley, P. Laloyaux, P. Lopez, C. Lupu, G. Radnoti, P. de Rosnay, I. Rozum, F. Vamborg, S. Villaume, J.-N. Thépaut, The ERA5 global reanalysis. *Quarterly Journal of the Royal Meteorological Society* **146**, 1999–2049 (2020).

47. OpenAI *et al.*, GPT-4 Technical Report. arXiv:2303.08774 [cs.CL] (2023).

48. D. Guo *et al.*, DeepSeek-R1 incentivizes reasoning in LLMs through reinforcement learning. *Nature* **645**, 633–638 (2025).

49. Chinese Academy of Meteorological Science, China Meteorological Administration, *Yearly charts of dryness/wetness in China for the last 500-year period* (Cartological Press, Beijing, China, 1981).

50. D. Zhang, *A Compendium of Chinese Meteorological Records of the Last 3000 Years* (Jiangsu Education House, Nanjing, 2004).

51. Standardization Administration of China, "Calculation and Promulgation of the Chinese Calendar" (GB/T 33661-2017, National Standard of the People's Republic of China, 2017).

52. H. Aslaksen, "The mathematics of the Chinese calendar" (National University of Singapore, Singapore, 2010).

53. P. von Platen, S. Patil, A. Lozhkov, P. Cuenca, N. Lambert, K. Rasul, M. Davaadorj, D. Nair, S. Paul, W. Berman, Y. Xu, S. Liu, T. Wolf, Diffusers: State-of-the-art diffusion models, version 0.34.0, Hugging Face (2022); https://github.com/huggingface/diffusers.

54. J. Ho, W. Chan, C. Saharia, J. Whang, R. Gao, A. Gritsenko, D. P. Kingma, B. Poole, M. Norouzi, D. J. Fleet, T. Salimans, Imagen video: High definition video generation with diffusion models. arXiv:2210.02303 [cs.CV] (2022).

55. S. Bassetti, B. Hutchinson, C. Tebaldi, B. Kravitz, DiffESM: Conditional emulation of temperature and precipitation in Earth system models with 3D diffusion models. *Journal of Advances in Modeling Earth Systems* **16**, e2023MS004194 (2024).

56. S. Rasp, P. D. Dueben, S. Scher, J. A. Weyn, S. Mouatadid, N. Thuerey, WeatherBench: a benchmark data set for data-driven weather forecasting. *Journal of Advances in Modeling Earth Systems* **12**, e2020MS002203 (2020).

57. S. Rasp, S. Hoyer, A. Merose, I. Langmore, P. Battaglia, T. Russell, A. Sanchez-Gonzalez, V. Yang, R. Carver, S. Agrawal, M. Chantry, Z. Ben Bouallegue, P. Dueben, C. Bromberg, J. Sisk, L. Barrington, A. Bell, F. Sha, WeatherBench 2: A benchmark for the next generation of data-driven global weather models. *Journal of Advances in Modeling Earth Systems* **16**, e2023MS004019 (2024).

58. National Oceanic and Atmospheric Administration, Oceanic Niño Index, National Centers for Environmental Information (2025); https://www.ncei.noaa.gov/access/monitoring/enso/sst.

59. O. Boucher *et al*., Presentation and evaluation of the IPSL-CM6A-LR climate model. *Journal of Advances in Modeling Earth Systems* **12**, e2019MS002010 (2020).

**Fig. S1.**
**Scheme of the calendar convention adopted in this study.** (A) Relationship between the standardized Gregorian and Chinese calendars. Both calendars are standardized to 12 months with 30 days per month. Since the Chinese New Year falls between January 21 and February 20 in the Gregorian calendar (corresponding to February 21 in the standardized solar calendar herein, given the 30-day constraint for January), two extreme scenarios for the earliest and latest Chinese New Year are illustrated. (B) Approximation of meteorological seasons and ancient Chinese seasonal divisions relative to the standardized Chinese calendar.

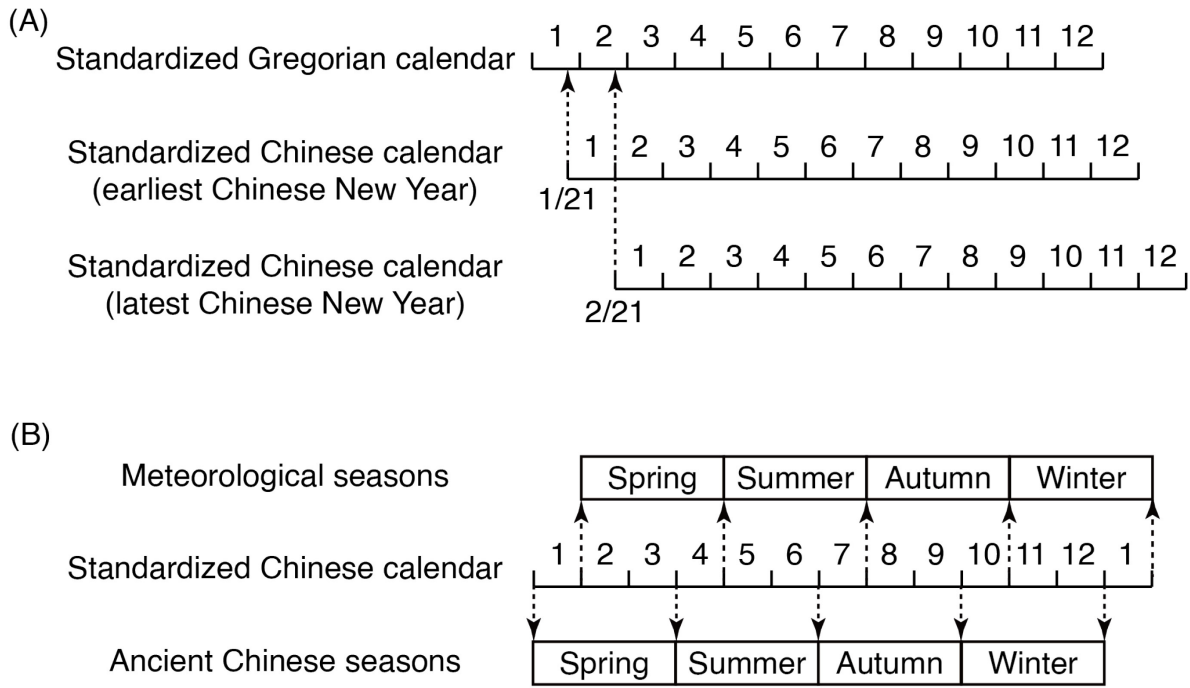

**Fig. S2.**
**Spatial distribution of documentary records.** Maps show the number of years containing recorded climate events at each grid point for (A) annual events (flood/drought) and for sub-annual events in (B) spring, (C) summer, (D) autumn, and (E) winter. The record density reflects historical population distribution, with higher counts in the long-settled cores of northern and eastern China and lower counts in the less densely populated southwestern region.

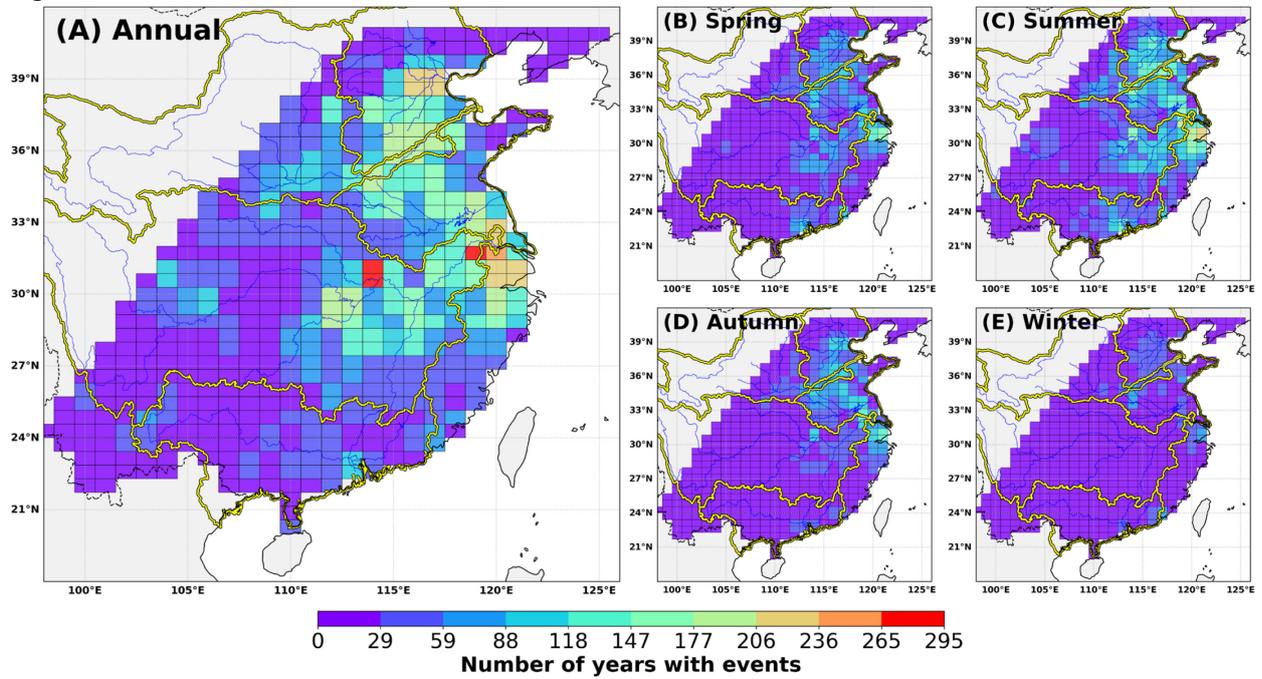

**Fig. S3.**

**Temporal coverage of documentary records.** Time series of the number of grid points containing recorded climate events for each year from 1368 to 1911. The series for annual events and for sub-annual events in each season are shown. Horizontal arrows on the right indicate the long-term average for each series. The data show a strong seasonal bias, with the most events recorded in summer and the fewest in winter. The lower event counts in the first ~100 years of the record are consistent with expected documentary loss over time. The total number of grid points in the study area is 526.

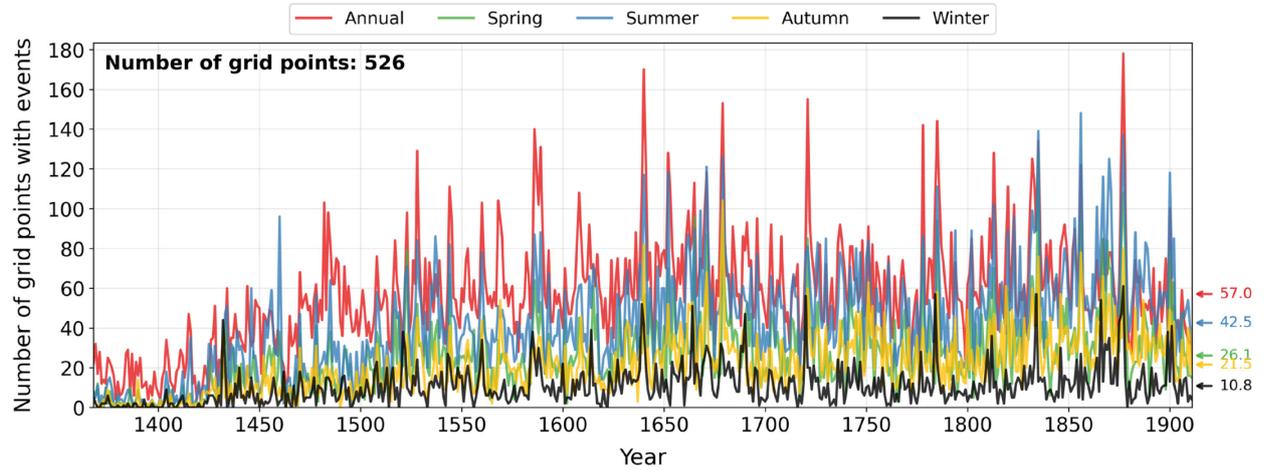

**Fig. S4.**
**Spatial distribution of the square of the Pearson correlation coefficient (RSQ) for annual and seasonal precipitation during validation**. RSQ values for seasonal precipitation are generally lower than those for annual precipitation due to sparser historical records and greater precipitation variability at finer timescales. For the same reason, our model exhibits superior performance in reconstructing summer precipitation compared to winter precipitation.

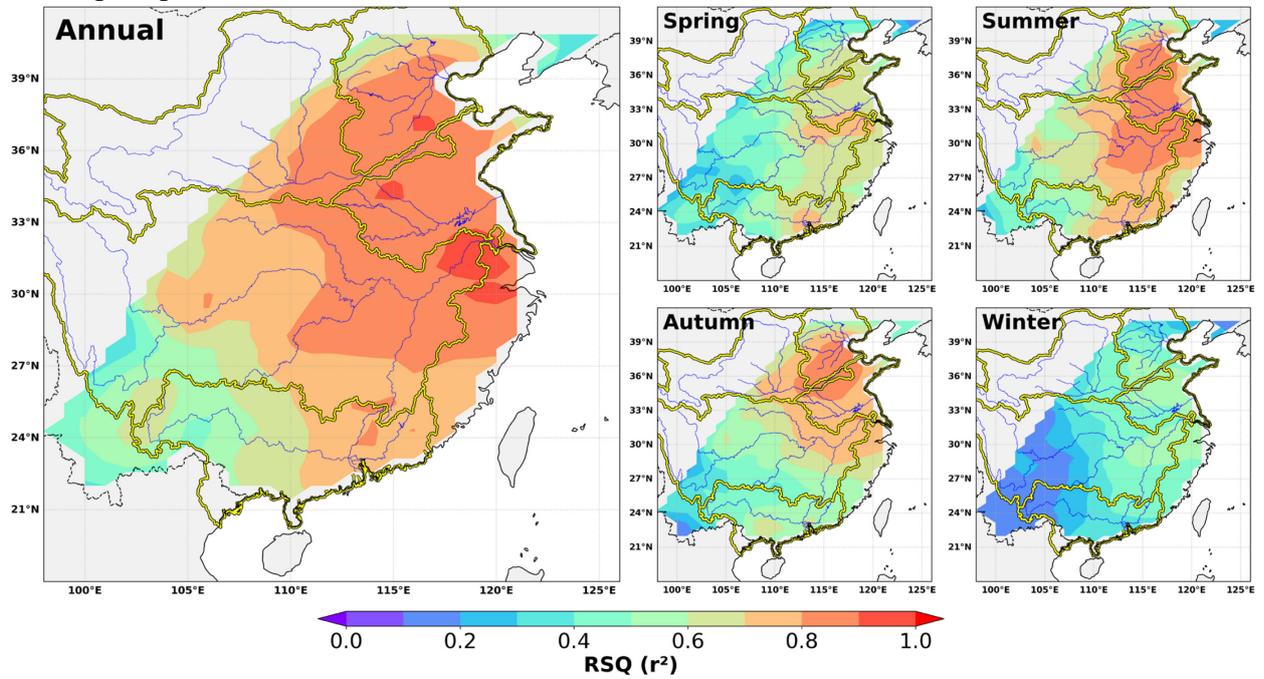

**Fig. S5.**
**Spatial distribution of the coefficient of efficiency (CE) for annual and seasonal precipitation during validation**. Consistent with RSQ, CE values for seasonal precipitation are generally lower than those for annual precipitation, reflecting sparser historical records and greater precipitation variability at finer temporal scales. Summer precipitation reconstruction outperforms winter precipitation for the same reason.

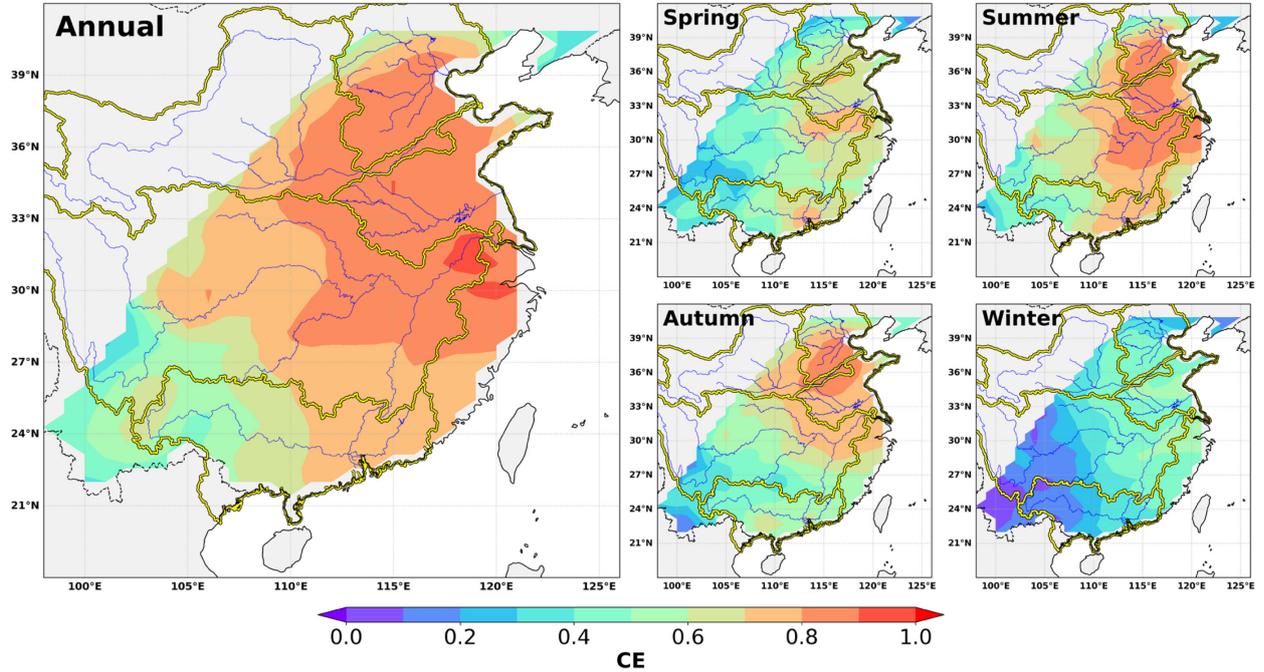

**Fig. S6.**
**Spatial distribution of the continuous ranked probability skill score (CRPSS) for annual and seasonal precipitation during validation**. Consistent with RSQ and CE, CRPSS values for seasonal precipitation are generally lower than those for annual precipitation, reflecting sparser historical records and greater precipitation variability at finer temporal scales. Summer precipitation reconstruction outperforms winter precipitation for the same reason.

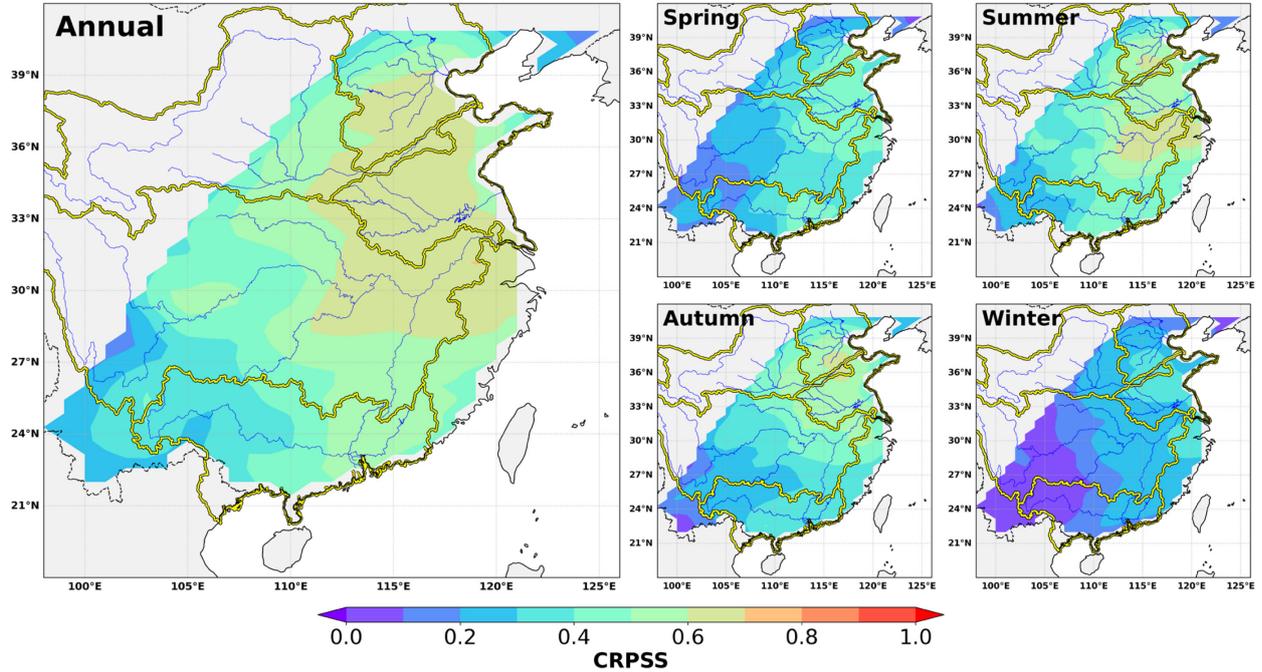

**Fig. S7.**
**Architecture of the 3D diffusion model backbone.** The model employs a symmetric encoder-decoder (4 blocks each, 4 layers per block) with cross-attention at the bottleneck. Conditioning on the annual progress fraction of the start date $(y, m, d)$ of the annual sequences $d_s = \frac{\text{DoY}(y,m,d)}{\text{DiY}(y)}$ (DoY = day of year, DiY = number of days in year) is critical, because it significantly expands the effective training data by allowing sequences to start on any calendar day, and it enables the model to account for the variable timing of the Chinese New Year in the Gregorian calendar. The architecture processes spatiotemporal tensors with dimensions: $N = 12$ (months), $C = 3$ (feature channels), and $(H, W) = (64, 64)$ (spatial grid).

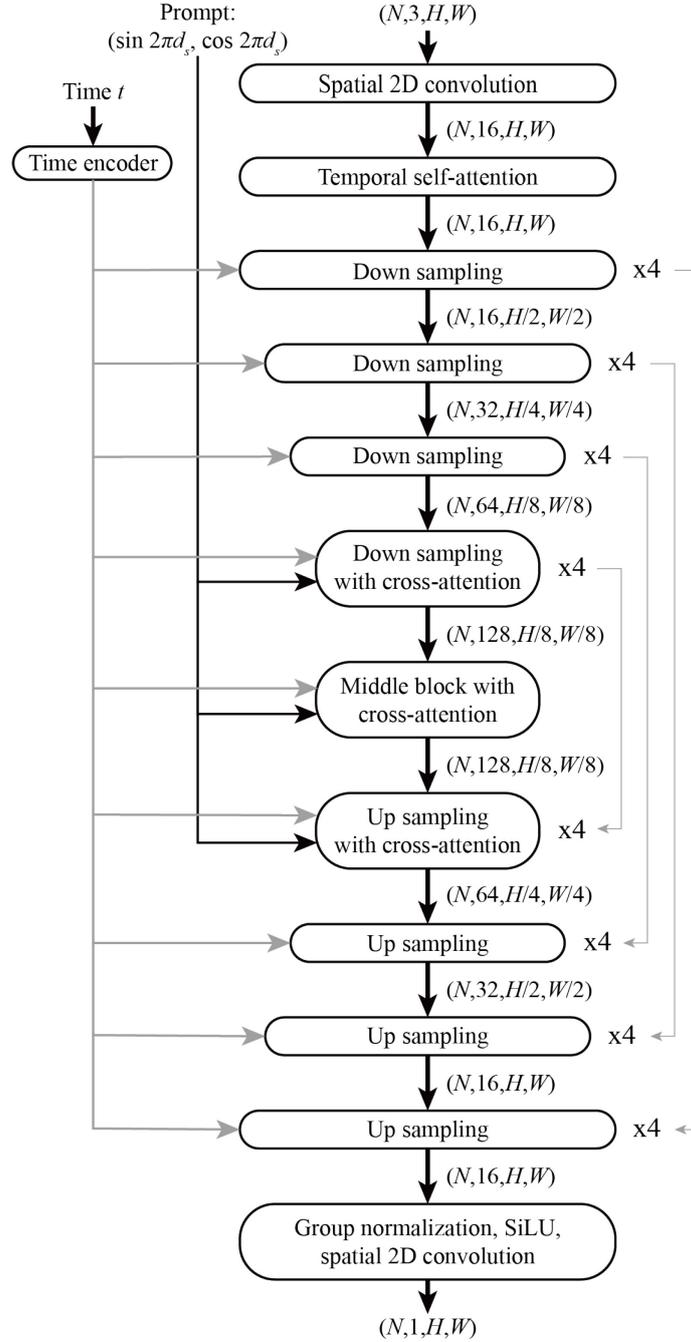

**Fig. S8.**
**Correlation between our reconstructed precipitation (May–September, MJJAS) and that of (*25*)**. The correlation exceeds 0.5 across most of the study area and reaches as high as 0.75 in the North China Plain, and lower correlations occur in the southwestern region where both (*25*) and our dataset lack sufficient data coverage.

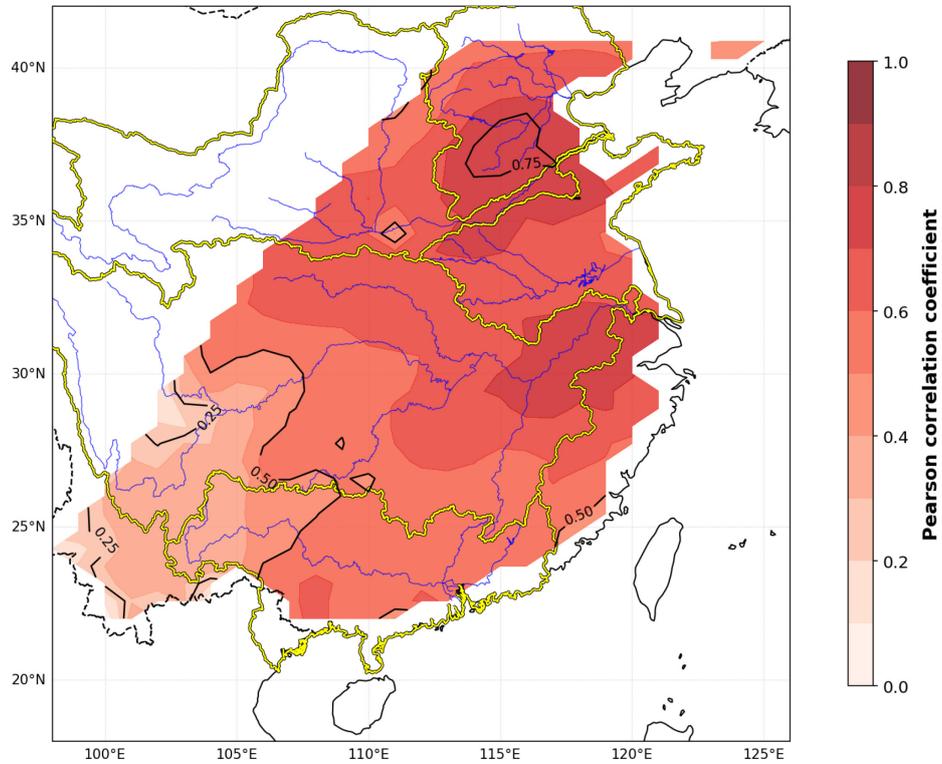

**Fig. S9.**
**Empirical Orthogonal Function (EOF) analysis of MJJAS precipitation from our reconstruction.** Comparison with Fig. 8 of (*25*) reveals good agreement in both the general spatial pattern and the boundaries between adjacent anomaly poles.

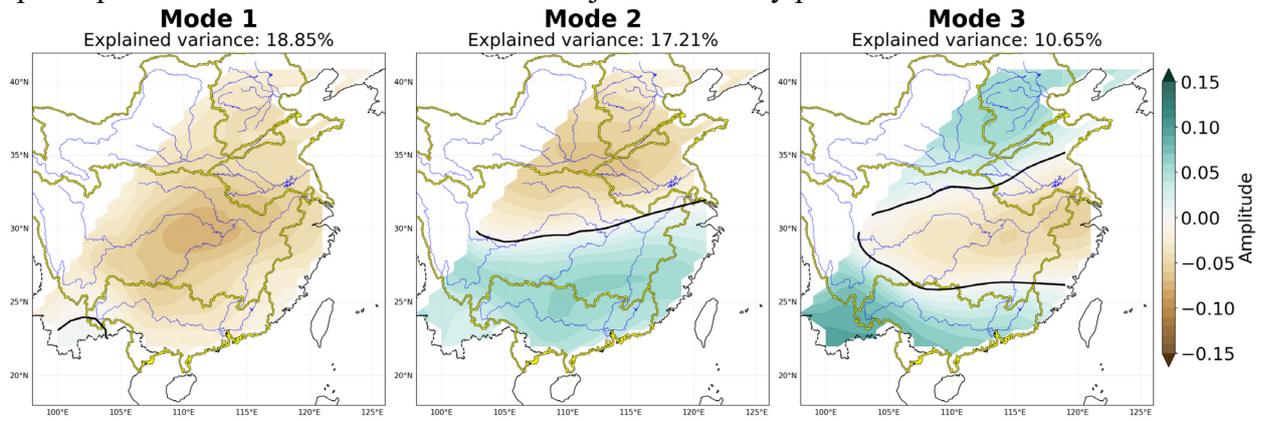

**Fig. S10.**

**The years with extreme principal component (PC) values corresponding to the first four EOF modes of the annual precipitation.** For each mode, the years with the highest and lowest PC values are listed in the top and bottom rows, respectively.

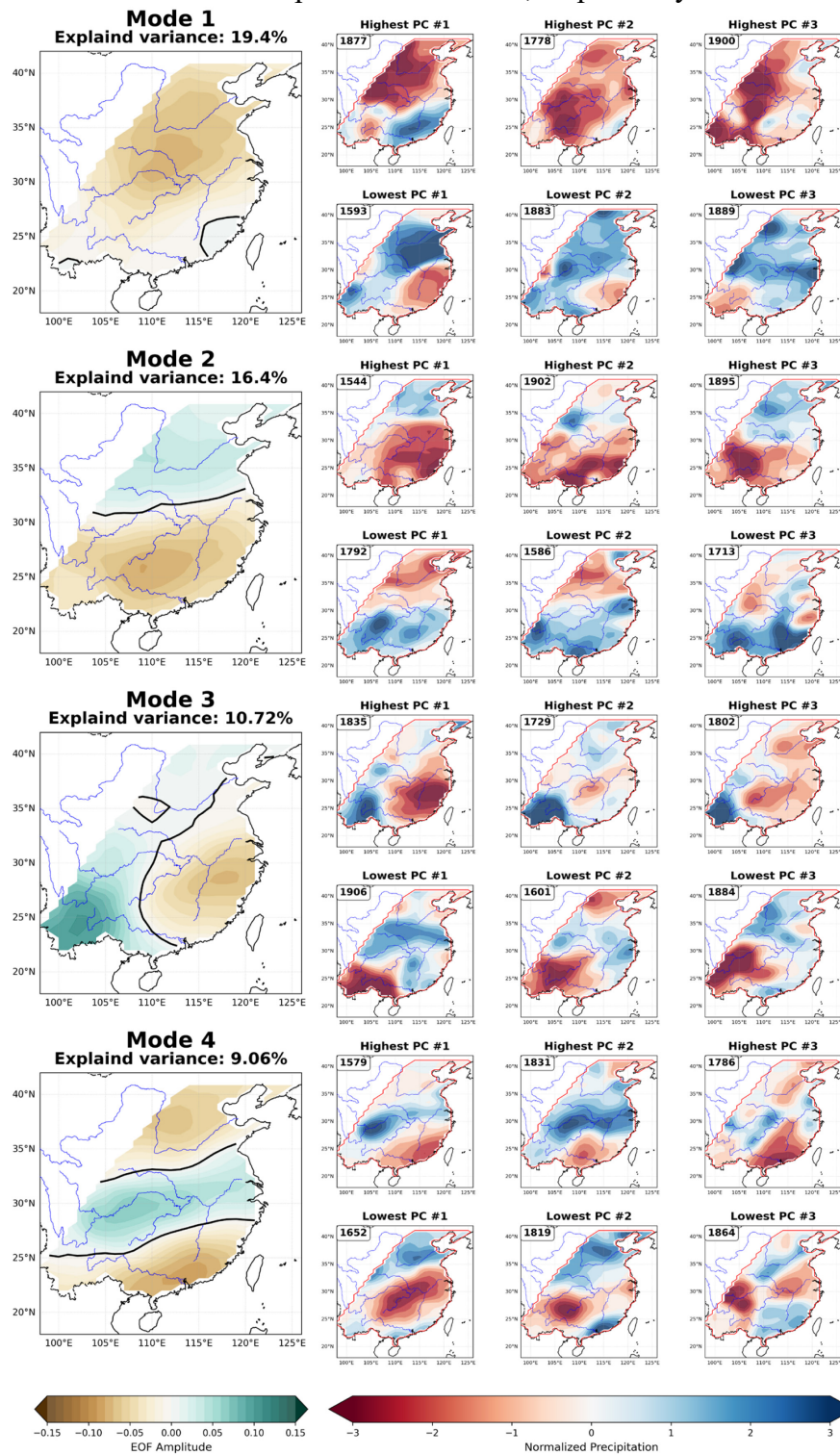

**Fig. S11.**
**EOF analysis of seasonal precipitation.** The two leading modes (Modes 1 and 2) correspond to a uniform spatial pattern and a meridional dipolar pattern in all seasons. The uniform mode precedes the meridional dipolar mode in summer, autumn, and winter, whereas the meridional dipolar mode precedes the uniform mode in spring. Modes 3 and 4 for all seasons are characterized by meridional tripolar or zonal bipolar patterns, which are analogous to the EOF modes derived from annual precipitation. Notably, the explained variances (EVs) of Modes 1 and 2 in winter are higher than those in the other seasons.

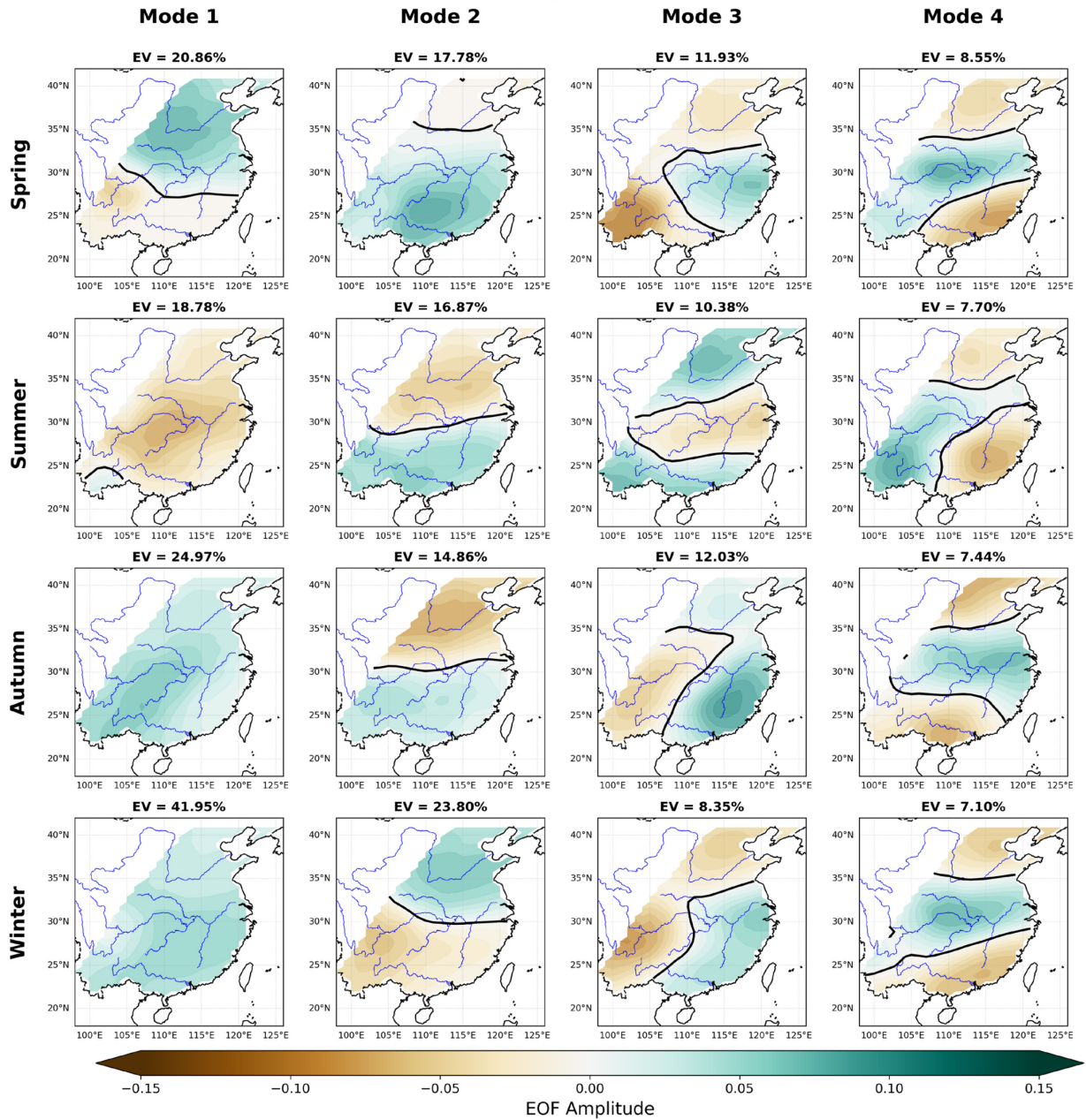

**Fig. S12.**

**Basin-averaged precipitation anomalies using a global normalization.** Time series (1368–1911) of normalized annual and seasonal precipitation, averaged within the five major river basins (Hai, Yellow, Huai, Yangtze, Pearl). Precipitation was normalized grid-point-wise using statistics from the full 1368–1911 period before averaged in the river basin. The top three flood (positive) and drought (negative) years for each basin are highlighted.

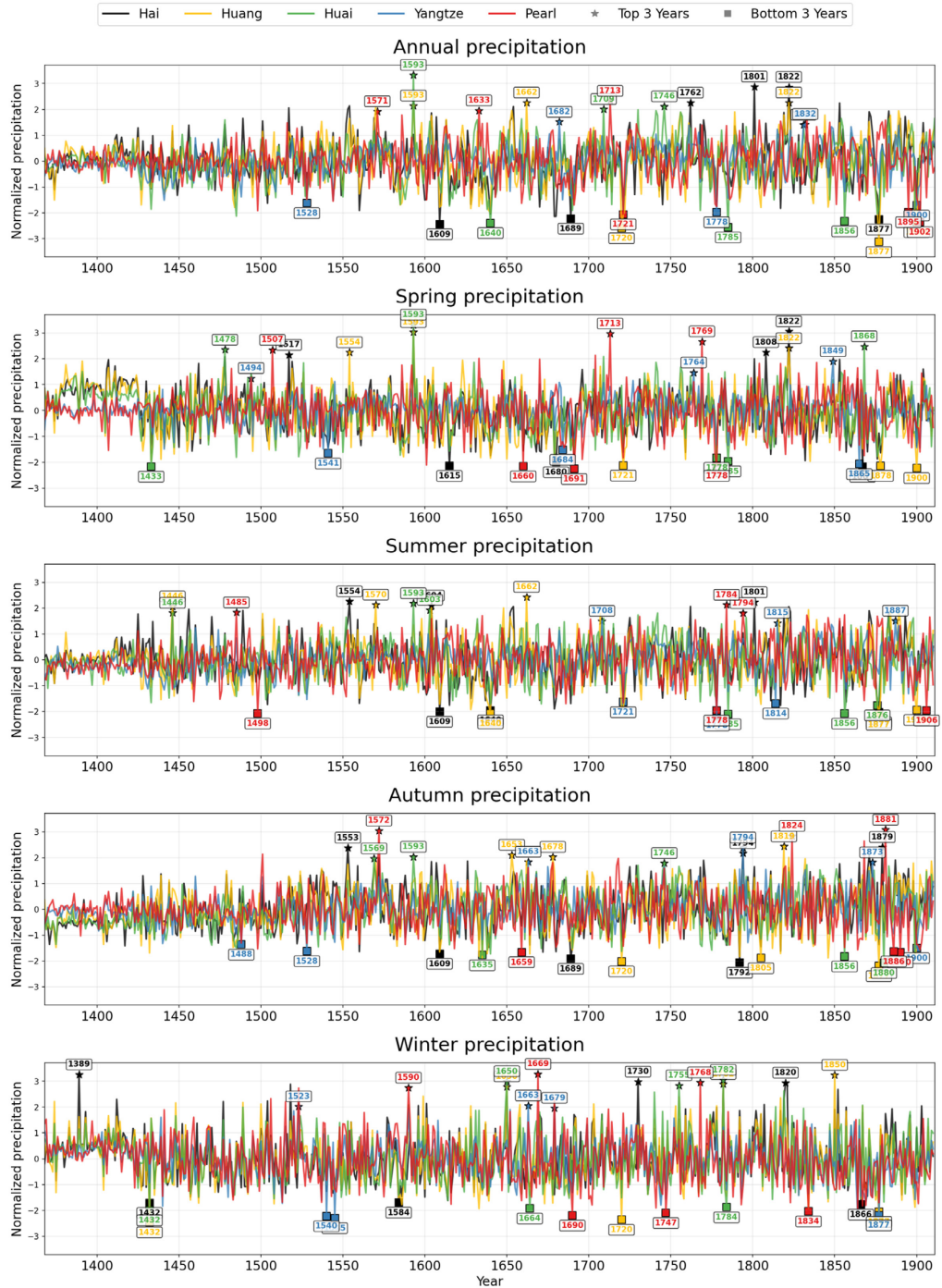

**Fig. S13.**

**Normalized annual and seasonal precipitation during the Chongzhen Drought (1637–1643)**. This drought is recognized as one of the most severe aridity events in the Ming and Qing Dynasties and has been identified as a critical contributing factor to the fall of the Ming Dynasty. The drought was most severe from the spring to autumn of 1640.

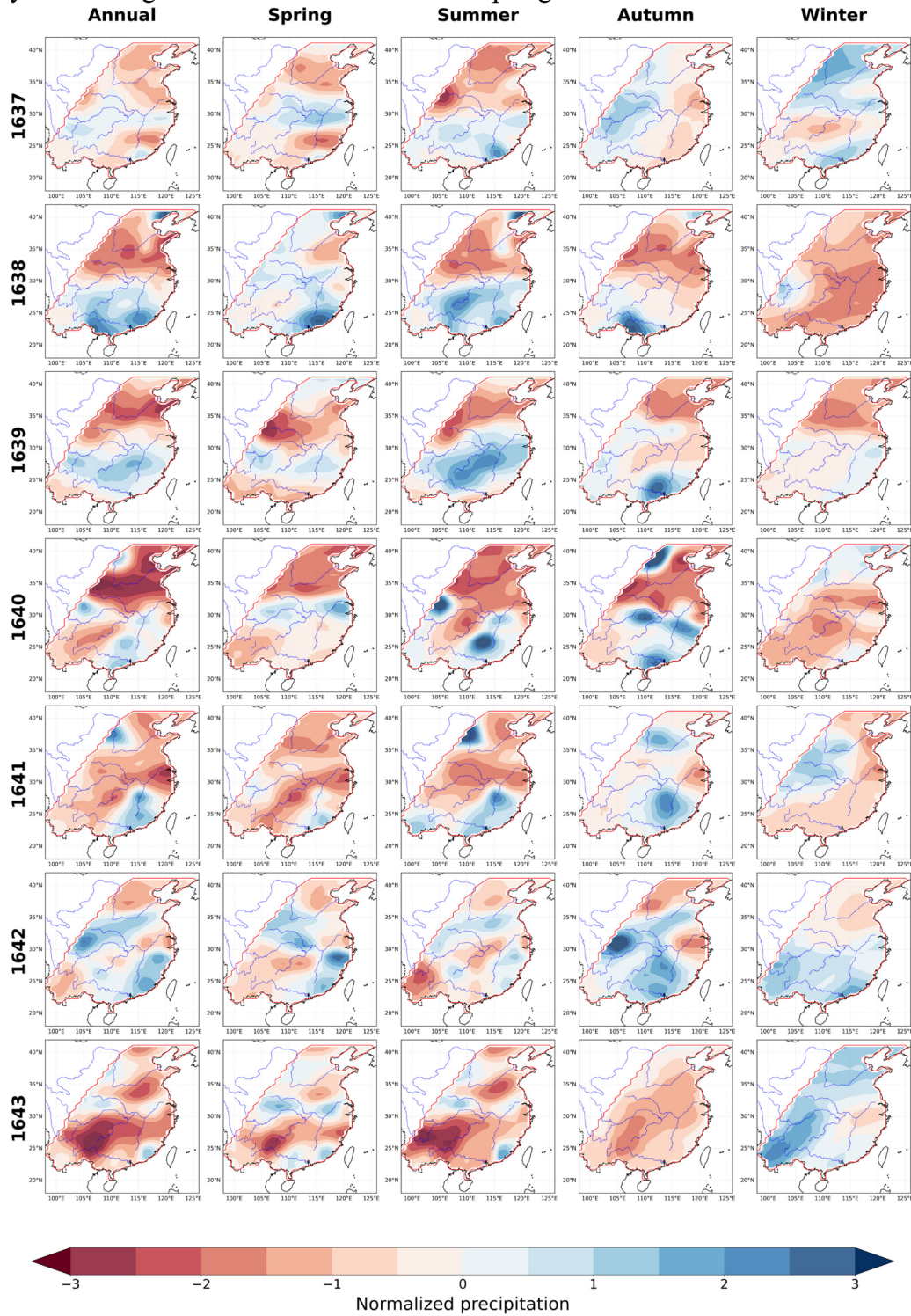

**Fig. S14.**

**Normalized seasonal precipitation averaged in China's five major river basins during the Chongzhen Drought (1637–1643).** During the most severe phase of this drought (spring to autumn of 1640), the Hai, Yellow, and Huai River Basins, which are all parts of northern China, exhibited normalized seasonal precipitation of approximately -2. While northern China experienced the most severe drought conditions overall, the drought in southern China (i.e., the Yangtze and Pearl River Basins) was also noteworthy in 1643.

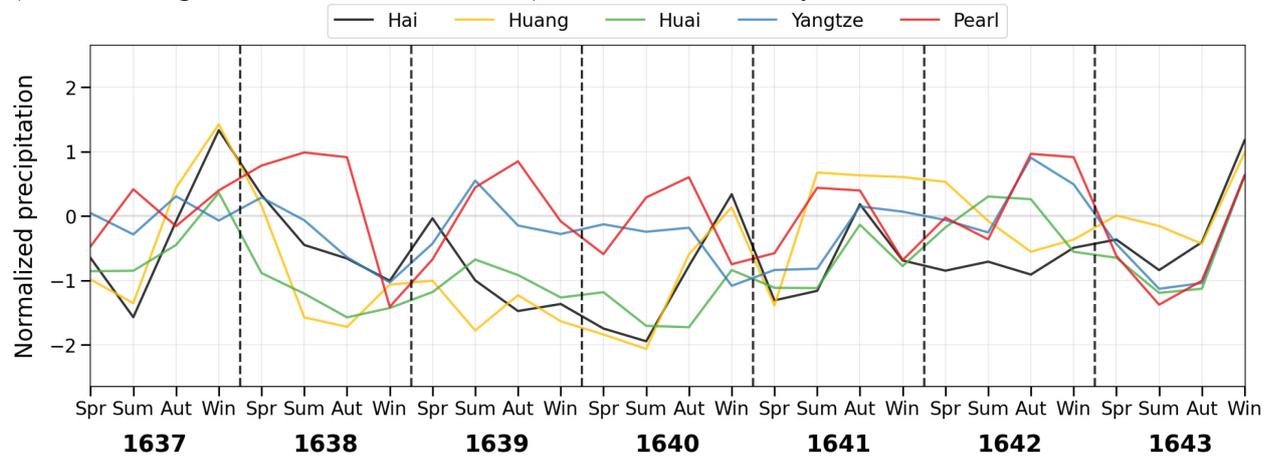

**Fig. S15.**

**Normalized annual and seasonal precipitation during the Guangxu Drought (1875–1879).** As a prominent aridity event in the Qing Dynasty, this drought caused over 10 million deaths in northern China. The most severe phase of the drought occurred from the spring of 1877 to the spring of 1878.

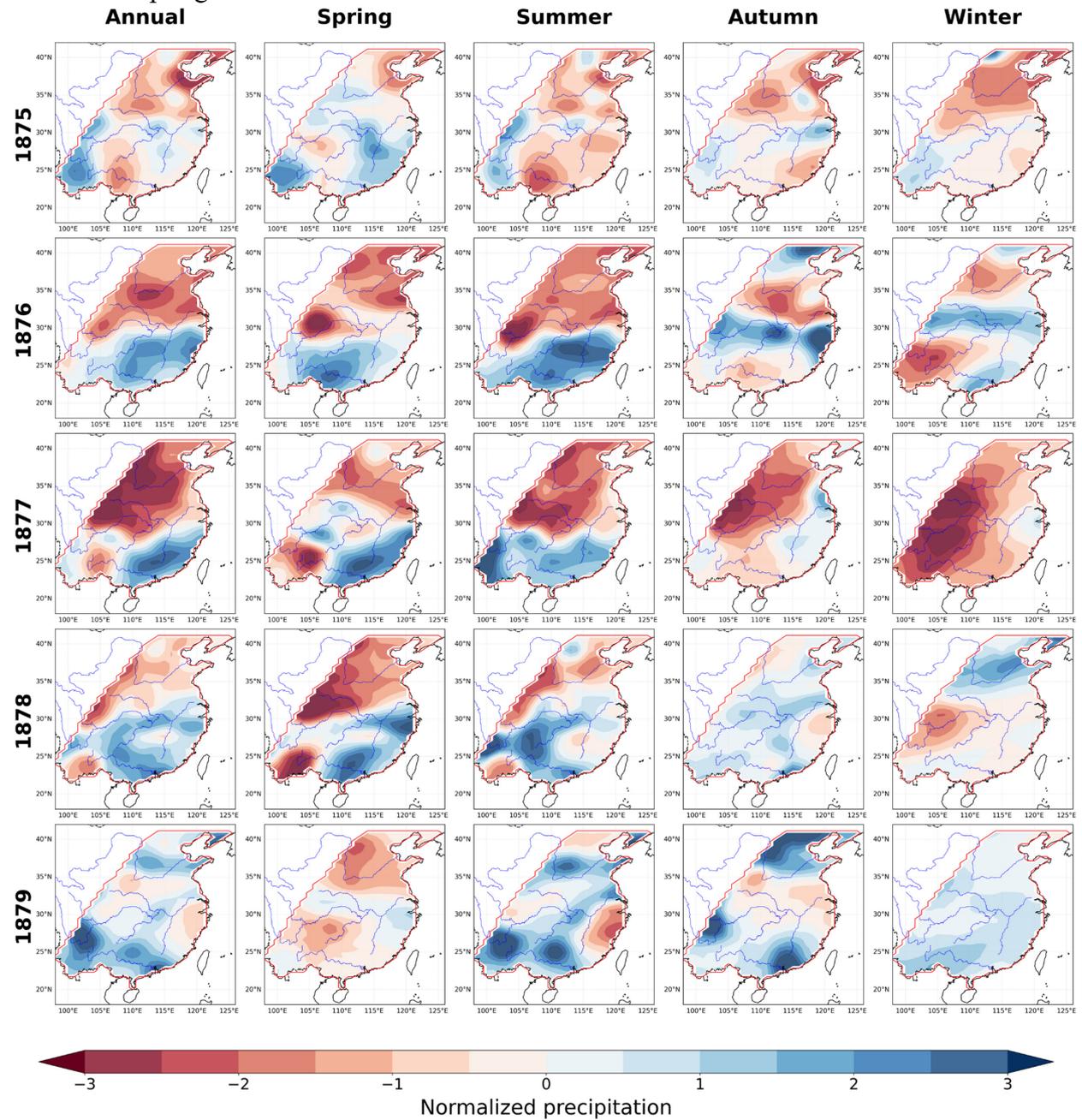

**Fig. S16.**

**Normalized seasonal precipitation averaged in China's five major river basins during the Guangxu Drought (1875–1879).** During the most severe phase of this drought (Spring 1877–Spring 1878), the Hai, Yellow, and Huai River Basins exhibited normalized seasonal precipitation of approximately -2, while the Yangtze and Pearl River Basins in southern China also experienced severe drought conditions around the winter of 1877.

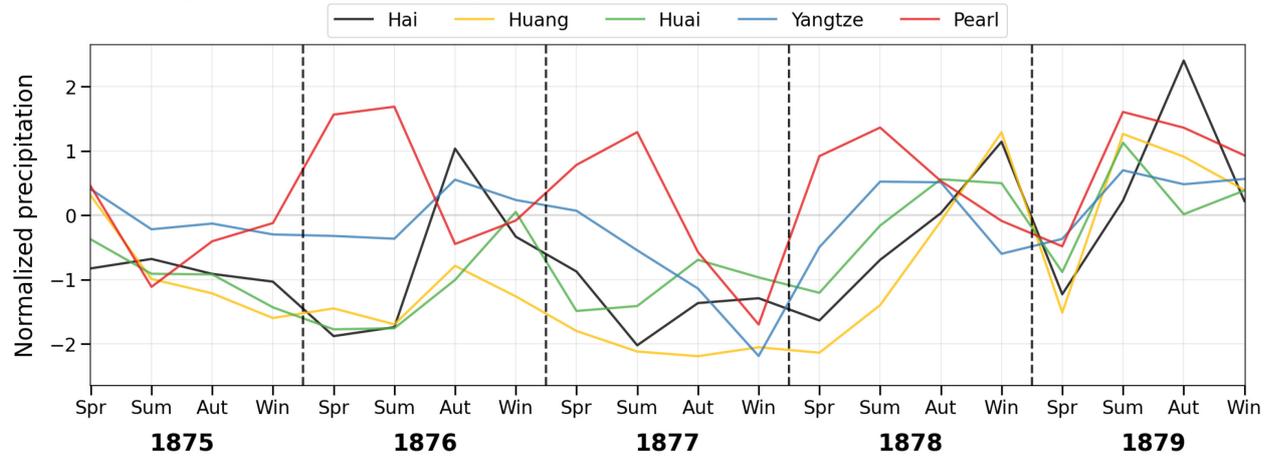

**Fig. S17.**

**Annual, seasonal, and monthly precipitation patterns for the year 1721.** This year witnessed a prominent widespread drought spanning northern, central, and southern China, with a notable monthly shift in the regions with precipitation exceeding the climatological mean. All the months referred to herein are lunar months. As discussed in the main text, these monthly values are suitable for qualitative and semi-quantitative reference but should be used with caution in high-precision quantitative applications.

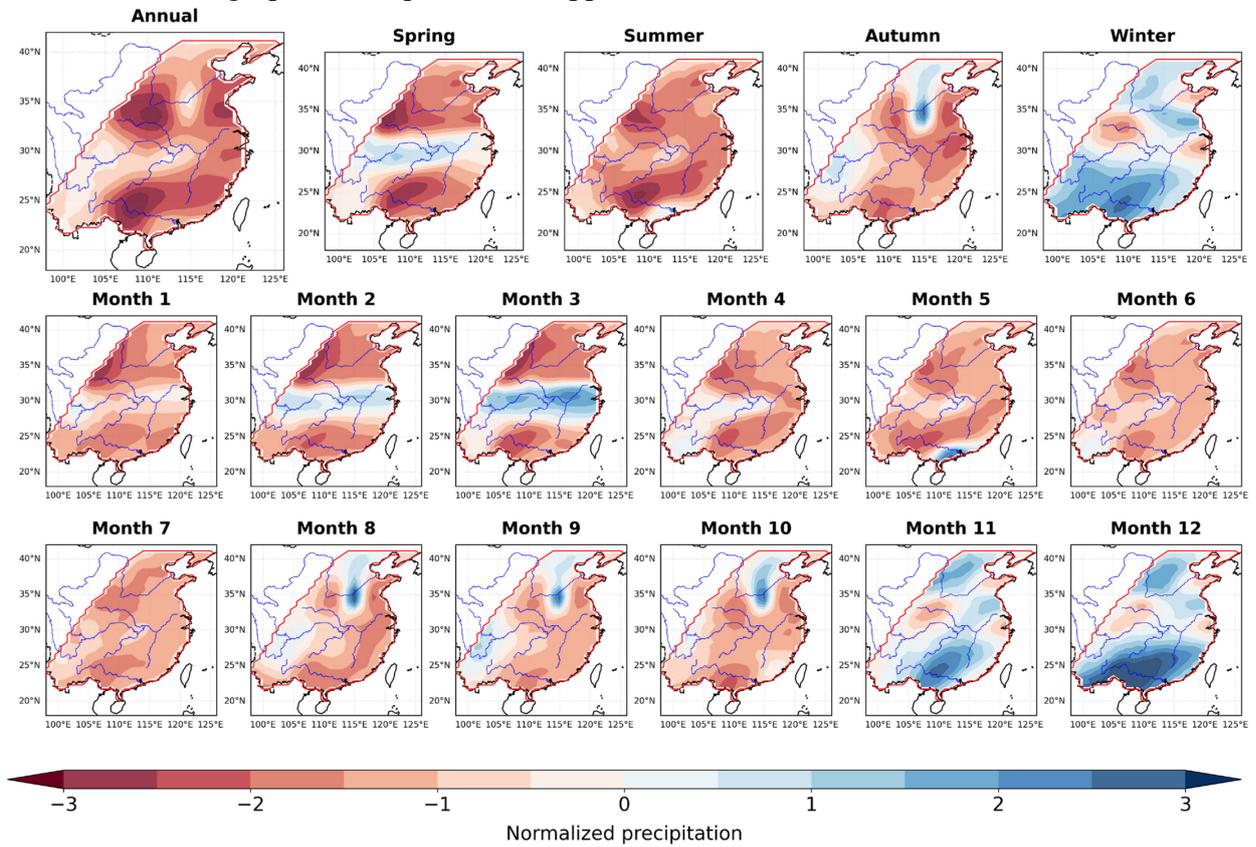

**Fig. S18.**

**Ablation study 1: Basin-averaged precipitation anomalies using IPSL-CM6A-LR piControl as training data.** Time series (1368–1911) of normalized annual and seasonal precipitation, processed identically to Fig. S12 but trained on a 1,930-year pre-industrial control (piControl) simulation from the IPSL-CM6A-LR model. The overall spatiotemporal patterns and anomaly amplitudes are consistent with the primary reconstruction (Fig. S12). The set of the most extreme years is largely conserved, with substantial overlap in the identified top flood and drought events for each basin. This demonstrates the reconstruction's stability to the choice of climate model used for training.

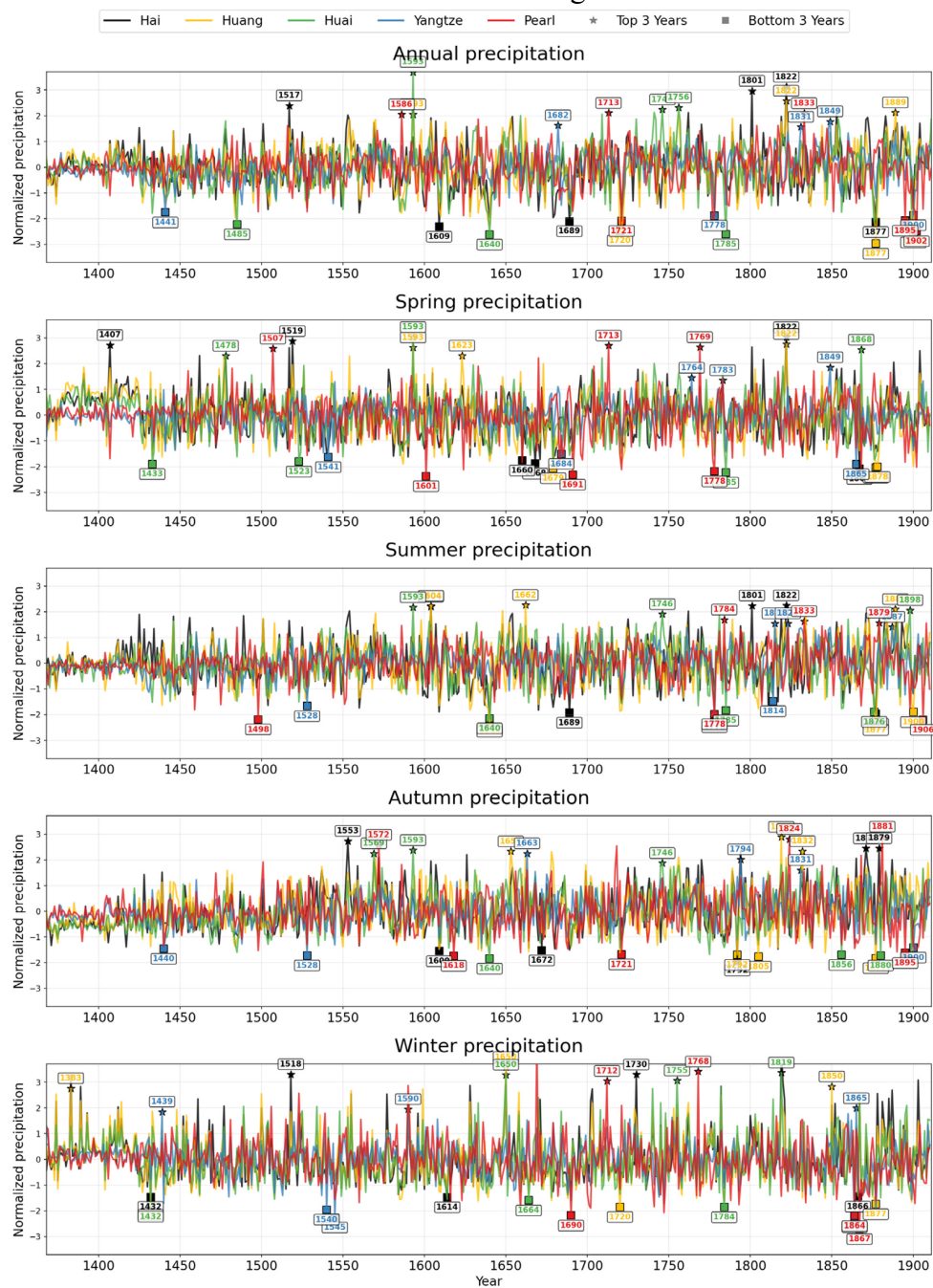

**Fig. S19.**

**Ablation study 1: Leading modes of climate variability using IPSL-CM6A-LR piControl as training data.** Empirical Orthogonal Function (EOF) analysis of annual precipitation (1368–1911). (A) The four leading modes from the reconstruction trained on the IPSL-CM6A-LR piControl simulation. (B) The corresponding modes from the IPSL-CM6A-LR piControl training data itself. The spatial patterns and explained variance (EV) of the leading modes in (A) are similar to those from our primary reconstruction (Fig. 4A), demonstrating robustness in the recovered large-scale climate modes. A detailed comparison reveals that the boundaries between poles in Modes 2 and 3 reflect characteristics present in the corresponding modes in the training data (B), illustrating how the model synthesizes information from the physical simulations.

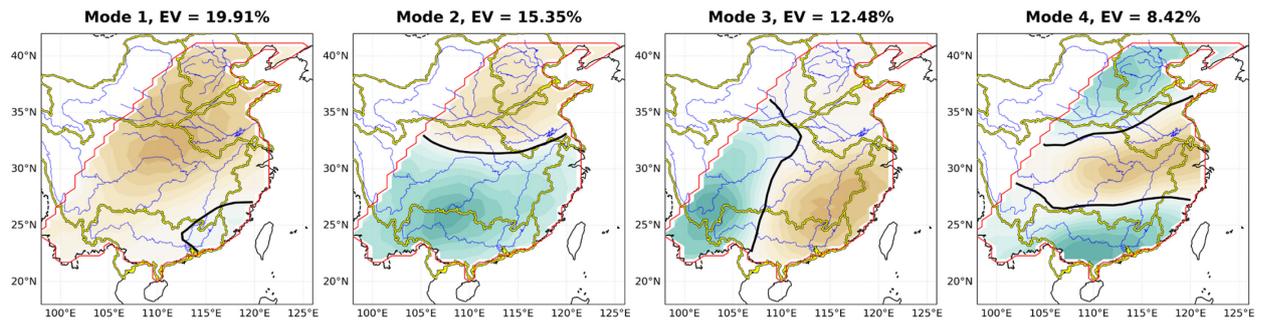
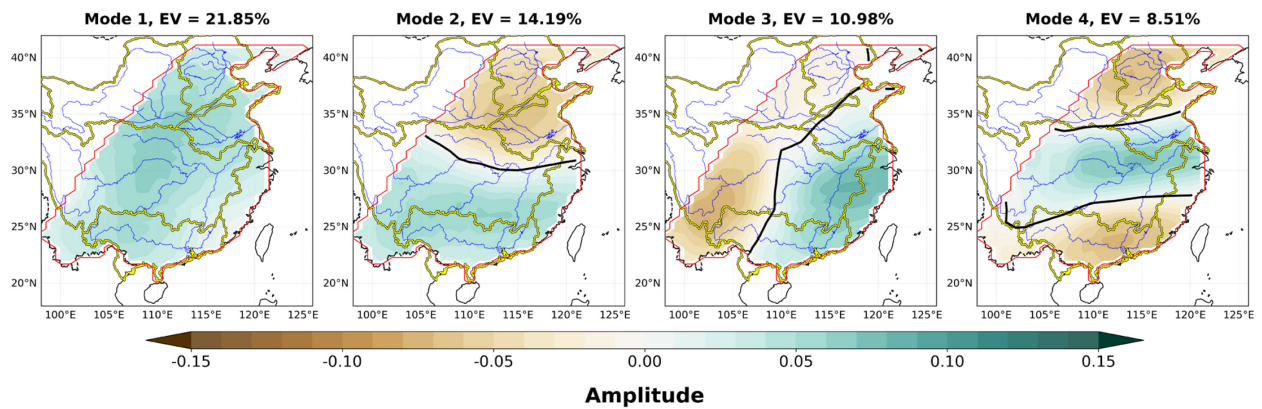

**Fig. S20.**

**Ablation study 1: ENSO-precipitation correlation using IPSL-CM6A-LR piControl as training data.** Spatial correlation between precipitation and the ENSO 3.4 index, where the reconstruction model is trained on the IPSL-CM6A-LR piControl simulation. Panels (A–E) show correlations for individual centuries (15th–19th), and Panel (F) shows the correlation for the full Ming–Qing period (1368–1911). The correlation patterns are consistent with those from our primary reconstruction (Fig. 4A–F), confirming the robustness of the identified multi-century teleconnection patterns to the choice of climate model used for training data.

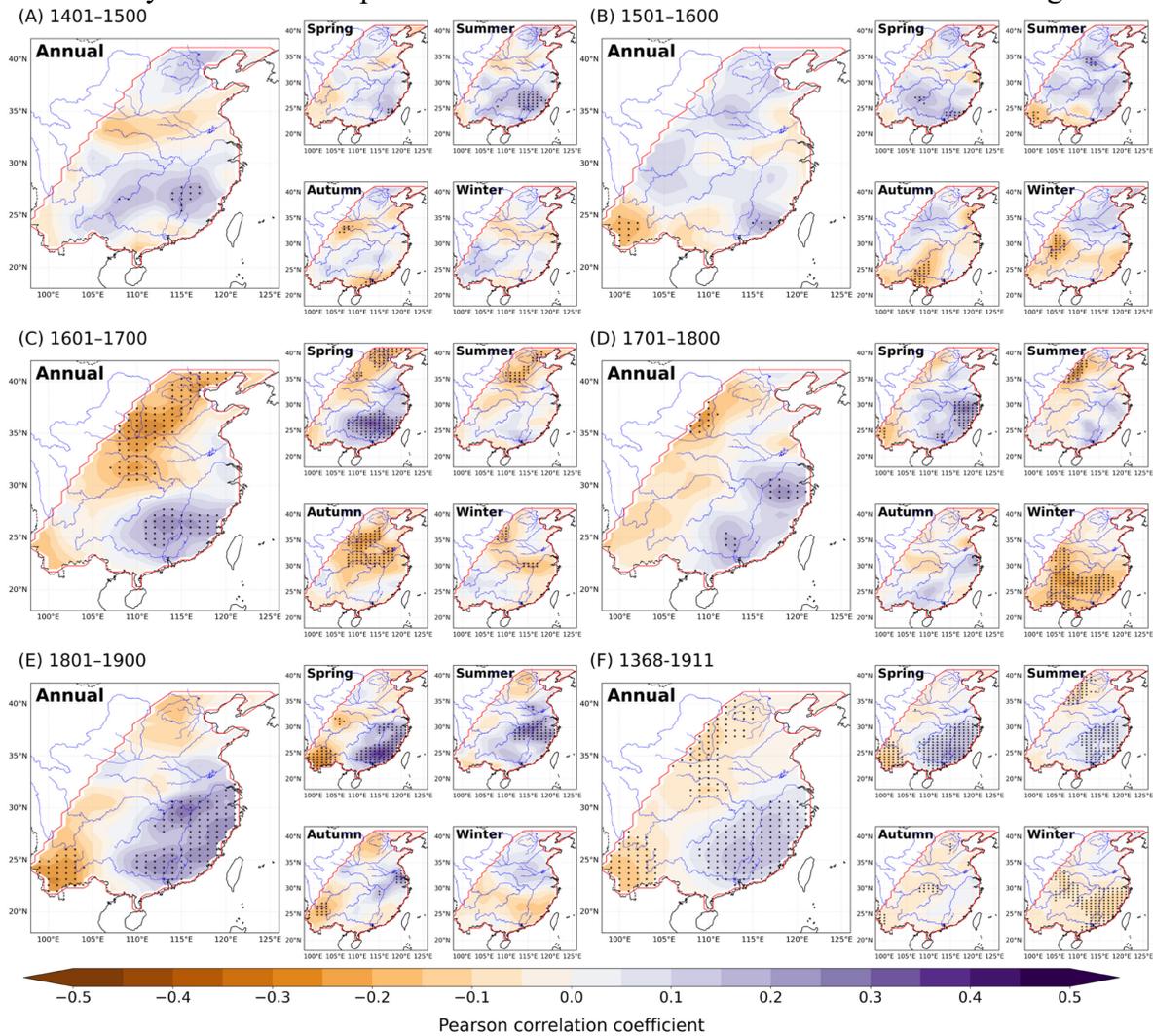

**Fig. S21.**

**Ablation study 2: Basin-averaged precipitation anomalies using a rank-based sliding window re-normalization.** Same as Fig. S12 but with precipitation normalized using a rank-based sliding window re-normalization (see Methods). Compared to Fig. S12, this method mitigates the dampening of early-record anomalies but reduces the relative intensity of the most extreme events across the full time series, illustrating a trade-off in re-normalization strategy.

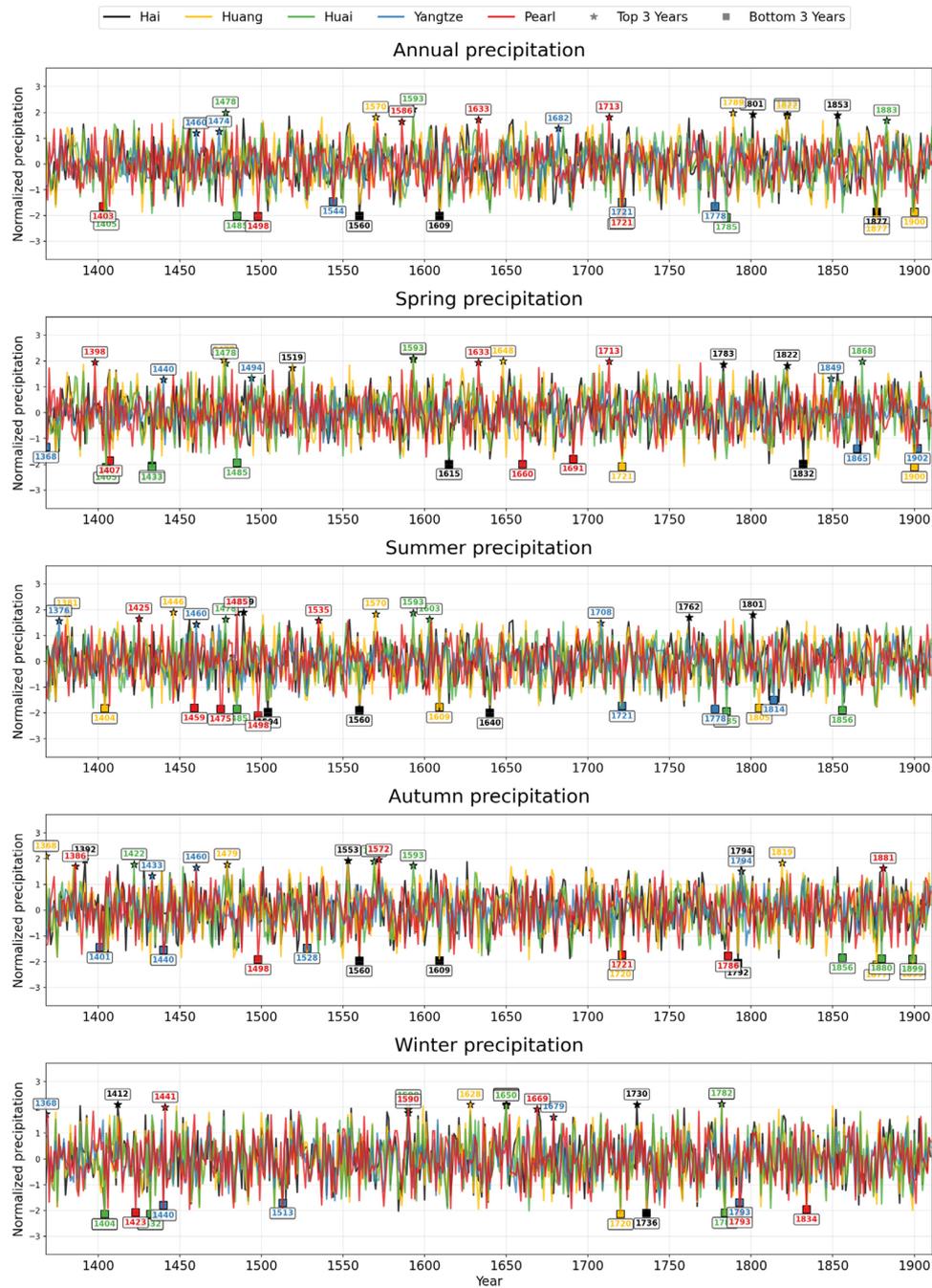

**Fig. S22.**

**Ablation study 2: Leading modes of climate variability using rank-based sliding window re-normalization.** Empirical Orthogonal Function (EOF) analysis of annual precipitation (1368–1911) normalized with the rank-based sliding window method. The spatial patterns of the four leading modes are similar to those derived from the globally normalized data (Fig. 3A). The primary difference is the order of the first two modes, which have nearly equal explained variance in both re-normalization schemes.

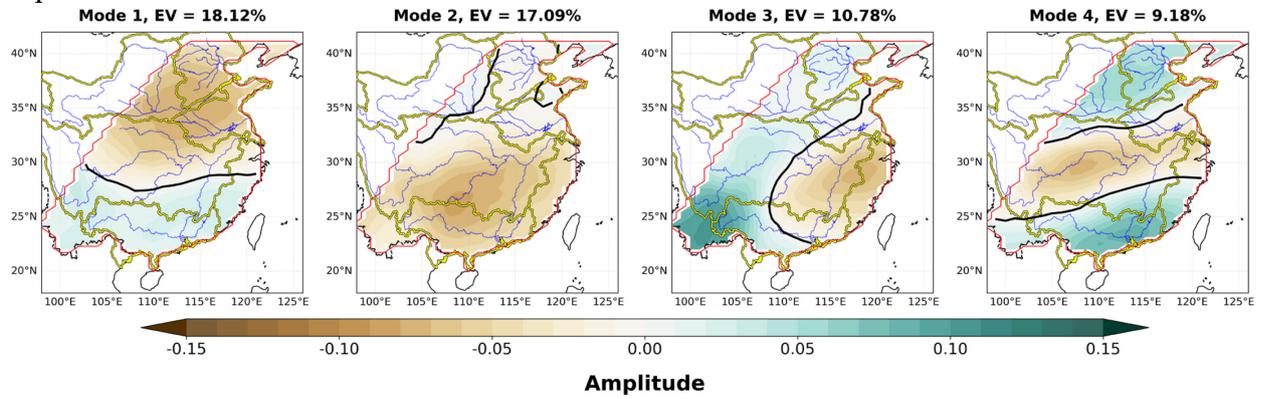

**Fig. S23.**

**Ablation study 2: ENSO-precipitation correlation using rank-based sliding window re-normalization.** Spatial correlation between precipitation and the ENSO 3.4 index, with precipitation normalized via the rank-based sliding window method. Panels (A–E) show correlations for individual centuries (15th–19th), and Panel (F) shows the correlation for the full Ming–Qing period (1368–1911). The correlation patterns are consistent with those derived using the global re-normalization method (Fig. 4A–F), confirming the robustness of the identified multi-century teleconnection patterns to this methodological choice.

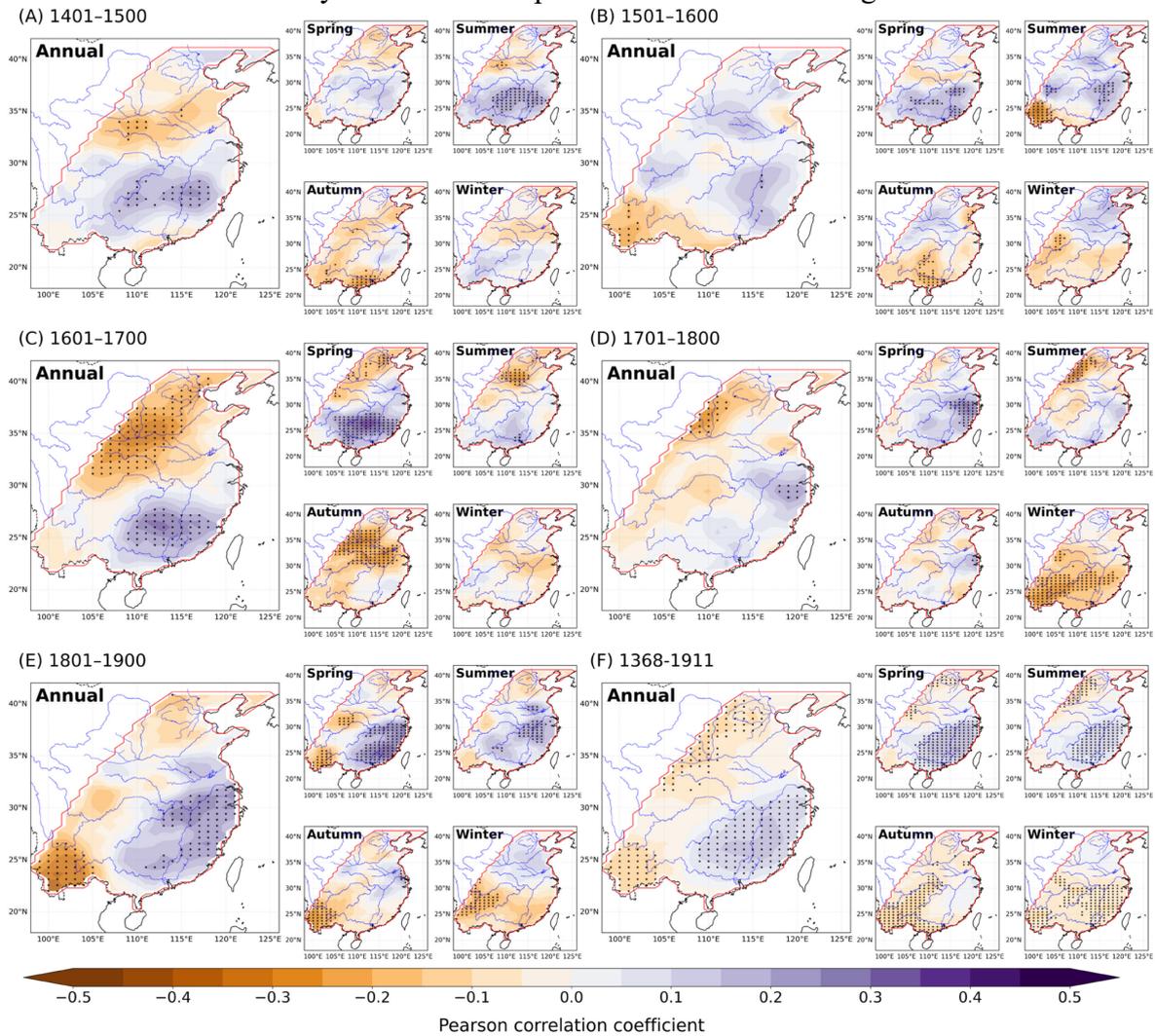
48

**Fig. S24.**

**Ablation study 3: Basin-averaged precipitation anomalies with an event dropout ratio of 0.2.** The overall fluctuation amplitude and the timing of the most extreme years (top floods/droughts, highlighted) are largely conserved compared to the full-record reconstruction (Fig. S12), demonstrating robustness to moderate gaps in the documentary archive.

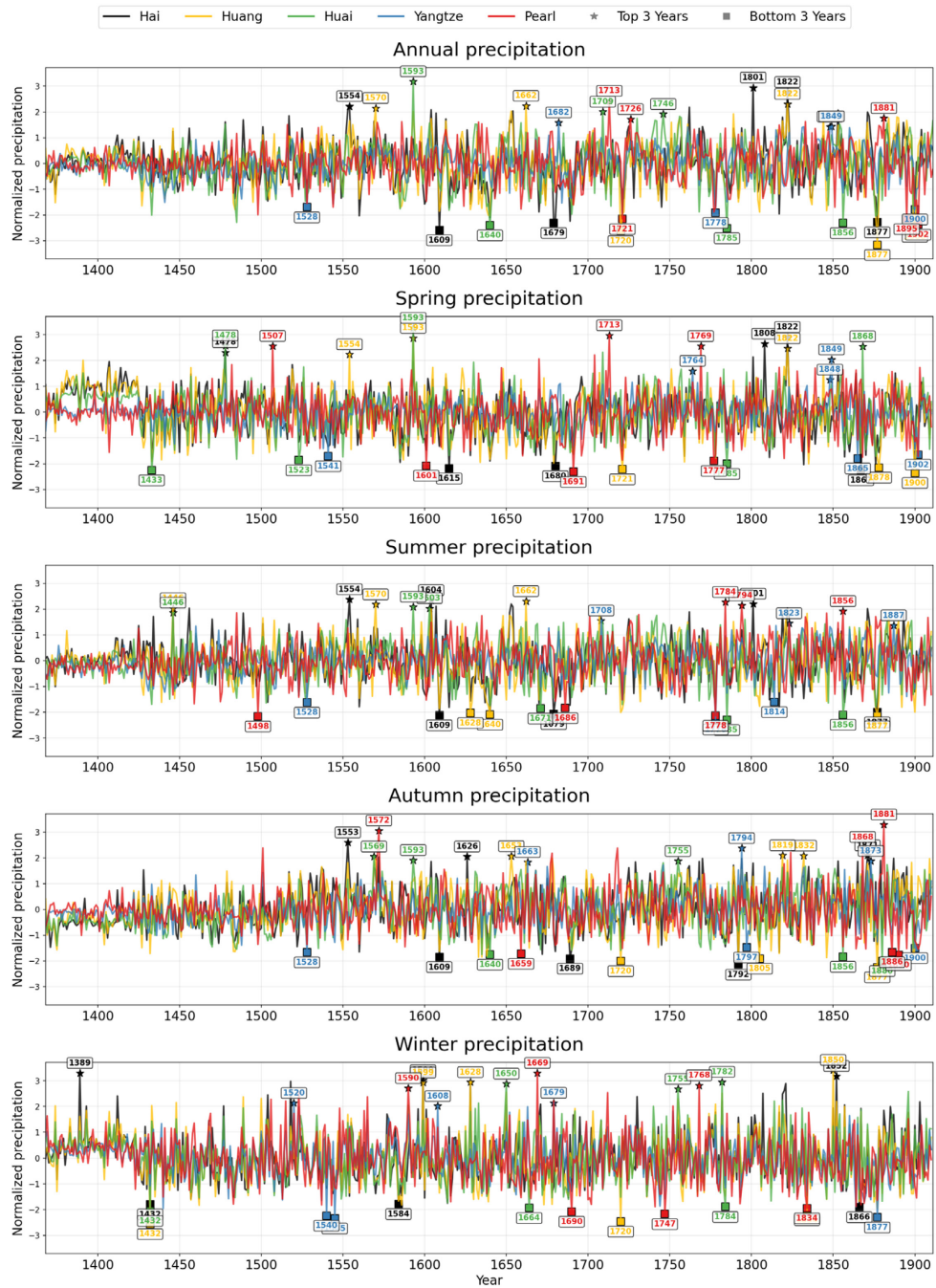

**Fig. S25.**

**Ablation study 3: Basin-averaged precipitation anomalies with an event dropout ratio of 0.5.** Despite the substantial reduction in input data, the reconstruction retains the core temporal evolution, anomaly amplitude, and identity of the most extreme years, indicating a high degree of resilience to even significant random record loss. Comparison with Figs. S12 and S23 shows consistent preservation of key climatic signals.

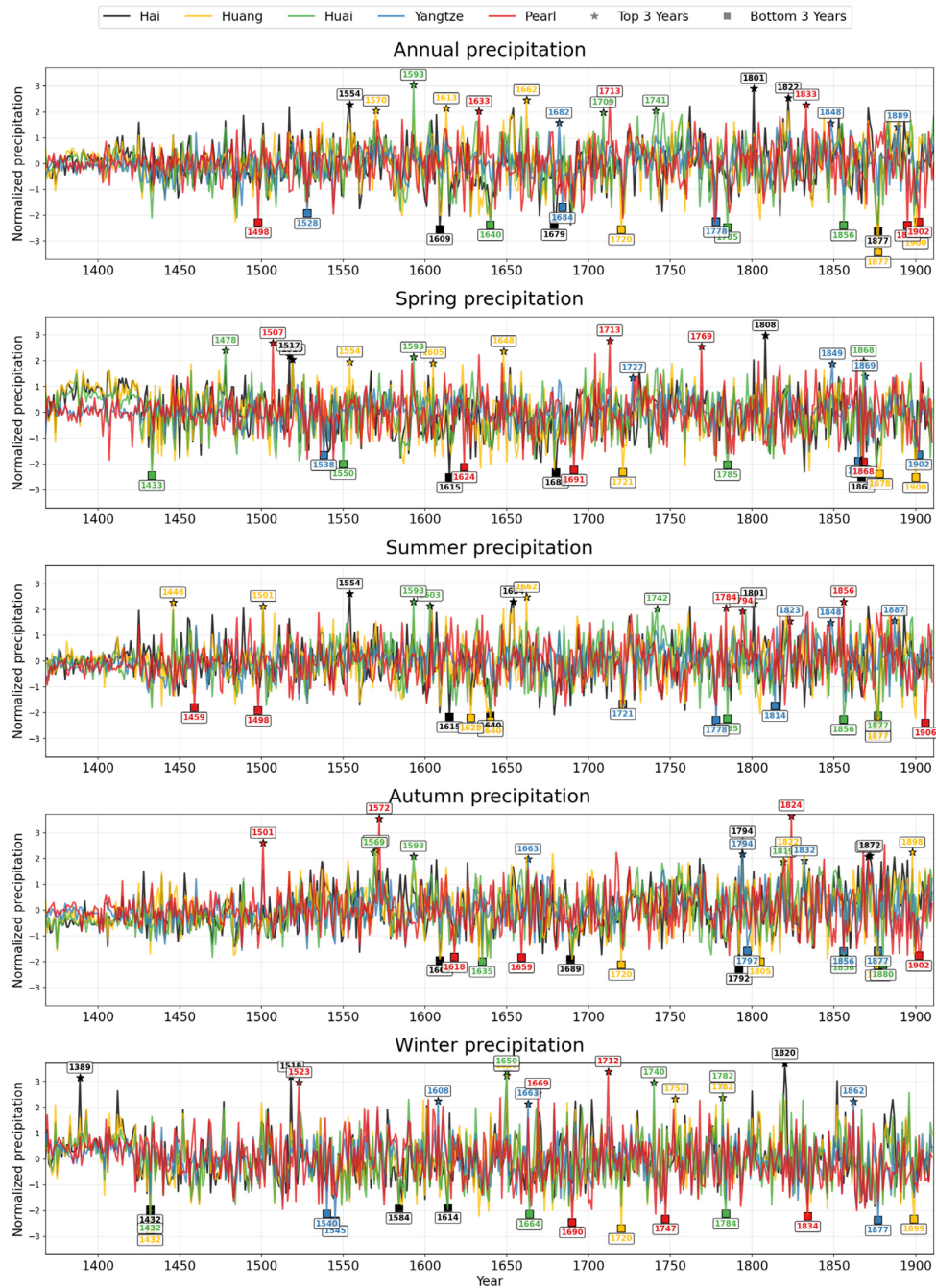

**Table S1.**

**Categorization of climate events in this study.** Each event type corresponds to a set of REACHES event codes (where "X" denotes a wildcard) and the associated time resolution. Annual flood and drought events are defined as those lacking sub-annual temporal information and are therefore resolved at the yearly level. All other events are categorized as sub-annual flood or drought. For the historical records specifying seasonal or multi-month timelines rather than precise monthly dates, we disaggregate them into individual months.

| Event name | REACHES event codes | Time resolution |
|---|---|---|
| Annual flood | 3101XXXXX, 3131XXXXX, 3141XXXXX, 3142XXXXX, 3151XXXXX | Annual |
| Annual drought | 10011XXXX, 100174XXX, 10111XXXX, 30XXXXXXX, 33XX101XX | Annual |
| Sub-annual flood | 3101XXXXX, 3131XXXXX, 3142XXXXX, 3151XXXXX | Sub-annual |
| Sub-annual drought | 10011XXXX, 100174XXX, 10111XXXX, 30XXXXXXX, 33XX101XX | Sub-annual |

**Table S2.**

**Examples of historical drought records and event categorization.** Historical drought records are presented with their original classical Chinese text, English translation, REACHES event codes, and our event categorization. The year and present-day location (in parentheses) are provided for context, as they are typically implied by the source document. All months refer to the Chinese calendar. Only REACHES event codes relevant to our drought categorization (Table S1) are listed.

| Original Chinese text | English translation | REACHES event code | Our event categorization |
|---|---|---|---|
| （1495年，陕西省陇县）岁罹大旱，雨泽愆期。今岁自春夏不雨，辫麦不收，秋禾未播。 | (Year 1495, Long County, Shaanxi Province) The year was struck by a severe drought, with rainfall arriving far behind its usual schedule. From spring to summer, no rain fell at all; barley yielded no harvest, and autumn crops could not even be sown. | 300130029 100174100 | Annual drought, sub-annual drought |
| （1536年，江苏省武进区）大旱。 | (Year 1536, Wujin District, Jiangsu Province) A severe drought. | 300130029 | Annual drought |
| （1640年，山西省襄垣县）春不雨，四月不雨，六月不雨。岁大饥，人相食。 | (Year 1640, Xiangyuan County, Shanxi Province) No rain fell in spring, no rain in the fourth month, no rain in the sixth month. The year saw a severe famine, and people resorted to cannibalism. | 100110189 | Sub-annual drought |
| （1801年，山东省荣成市）夏秋大旱，草木尽枯。冬饥。 | (Year 1801, Rongcheng City, Shandong Province) There was a severe drought in summer and autumn, with all plants and trees withering away. A famine then struck in winter. | 300130029 330110109 | Sub-annual drought |
| （1851年，山东省安丘县）诏免逋租。冬无雪。 | (Year 1851, Anqiu County, Shandong Province) An imperial edict was issued to exempt the overdue taxes. There was no snow at all in winter. | 101110089 | Sub-annual drought |
| （1877年，四川省大竹县）大旱，自四月至八月，无雨。 | (Year 1877, Dazhu County, Sichuan Province) A severe drought struck, with no rainfall recorded from the fourth to the eighth month. | 300130029 100110089 | Annual drought, sub-annual drought |

**Table S3.**

**Examples of historical flood records and event categorization.** Historical flood records are presented with their original classical Chinese text, English translation, REACHES event codes, and our event categorization. Formatting and conventions follow those of Table S2.

| Original Chinese text | English translation | REACHES event code | Our event categorization |
|---|---|---|---|
| （1376年，江苏省昆山市）今夏霪雨，又山水奔注，江湖增涨。 | (Year 1376, Kunshan City, Jiangsu Province) There was excessive and prolonged rain this summer; additionally, mountain torrents surged down, causing rivers and lakes to rise sharply. | 314130029 313100209 315100209 | Sub-annual flood |
| （1484年，福建省罗源县）霪雨，大水。 | (Year 1484, Luoyuan County, Fujian Province) Persistent rain and great flood. | 100150500 310130029 | Annual flood |
| （1593年，河南省西华县）夏五月，大雨潦麦。秋霖，复伤庄稼，县城四隅积水如湖，僻巷之隘者积水深二三尺不等。民家有网者出户便可得鱼。 | (Year 1593, Xihua County, Henan Province) In the fifth month, which falls in summer, a heavy downpour submerged the wheat fields. In autumn, continuous rainfall further damaged the crops. Water accumulated in all four corners of the county town, turning them into lakes; in the narrow alleys of remote areas, the water depth ranged from two to three feet. People who owned fishing nets could catch fish right outside their homes. | 310100309 310130029 310100009 310135029 | Sub-annual flood |
| （1734年，河南省原阳县）河溢，大堤以南水深数丈。 | (Year 1734, Yuanyang County, Henan Province) The river overflowed, and the water depth south of the main dyke reached several *zhang* (1 *zhang* ≈ 3.3 meters). | 315100209 315135029 | Annual flood |
| （1808年，安徽省潜山市）五月，潜山发蛟，水入淮，寿州长水二丈余，沿淮被淹。 | (Year 1808, Qianshan City, Anhui Province) In the fifth lunar month, a flash flood broke out in Qianshan, which flowed into the Huai River. The water level in Shouzhou rose by more than two *zhang*, and all areas along the Huai River were submerged. | 314230029 310135029 | Sub-annual flood |
| （1905年，广东省仁化县）城口大水，倒塌房屋一百八十余间，七星桥亦被冲断，县城高涨至丈六七尺。 | (Year 1905, Renhua County, Guangdong Province) Severe flooding hit Chengkou Town, with over 180 houses collapsing; the Qixing Bridge was also washed away, and the water level in the county town rose to one *zhang* and six to seven *chi* (1 *chi* ≈ 0.33 meter) deep. | 310130029 | Annual flood |

**Table S4.**

**Conversion table for meteorological seasons in the calendar conversion framework of this study.** Note that the Chinese New Year varies annually over the period from January 21 to February 20.

| Meteorological season | Gregorian months | Day range in standardized Chinese calendar (days after Chinese New Year) |
|---|---|---|
| Spring | March–May | 30–119 |
| Summer | June–August | 120–209 |
| Autumn | September–November | 210–299 |
| Winter | December–February | 300–389 |

**Table S5.**

**Variation of validation metrics during model training.** RSQ for annual, seasonal, and monthly precipitations and CRPSS for monthly precipitation are used to assess training convergence. Higher values of all metrics indicate better model performance. Based on these validation results, the model trained at Epoch 20 is selected as the final model.

| Epoch | RSQ | | | CRPSS | | |
|---|---|---|---|---|---|---|
| | Monthly precipitation | Seasonal precipitation | Annual precipitation | Monthly precipitation | Seasonal precipitation | Annual precipitation |
| 4 | 0.3812 | 0.4285 | 0.5664 | 0.2607 | 0.2686 | 0.4095 |
| 8 | 0.3957 | 0.4596 | 0.6138 | 0.2997 | 0.3195 | 0.4484 |
| 12 | 0.4007 | 0.4672 | 0.6397 | 0.3028 | 0.3285 | 0.4801 |
| 16 | 0.4009 | 0.4670 | 0.6402 | 0.3021 | 0.3290 | 0.4883 |
| 20 | 0.3993 | 0.4686 | 0.6486 | 0.2999 | 0.3279 | 0.4930 |

**Table S6.**

**Summary of validation statistics for precipitation reconstruction.** The table presents four key validation metrics: RSQ and CE for deterministic predictions, and CRPSS and SSR for ensemble predictions. Statistics are reported for annual precipitation, seasonal precipitation (spring, summer, autumn, winter, and their average), and monthly precipitation (January–December and their average). Higher RSQ, CE, and CRPSS values and SSR values closer to 1 indicate better model performance.

|  |  | RSQ | CE | CRPSS | SSR |
|---|---|---|---|---|---|
| Annual precipitation | | 0.6486 | 0.7278 | 0.6486 | 1.0328 |
| Seasonal precipitation | Spring | 0.4730 | 0.5202 | 0.3129 | 0.9642 |
| | Summer | 0.5968 | 0.6628 | 0.4328 | 0.9772 |
| | Autumn | 0.4933 | 0.5878 | 0.3722 | 0.9463 |
| | Winter | 0.3114 | 0.3336 | 0.1939 | 0.9000 |
| | **Average** | 0.4686 | 0.5261 | 0.3279 | 0.9469 |
| Monthly precipitation | January | 0.2440 | 0.3205 | 0.1847 | 1.0042 |
| | February | 0.2856 | 0.3633 | 0.2239 | 0.9151 |
| | March | 0.3230 | 0.3643 | 0.2225 | 0.9141 |
| | April | 0.3524 | 0.4111 | 0.2506 | 0.9006 |
| | May | 0.4703 | 0.5412 | 0.3412 | 0.9221 |
| | June | 0.5305 | 0.6005 | 0.3925 | 0.9192 |
| | July | 0.5583 | 0.6356 | 0.4194 | 0.9165 |
| | August | 0.5190 | 0.6280 | 0.4032 | 0.9387 |
| | September | 0.4948 | 0.5785 | 0.3677 | 0.9385 |
| | October | 0.4263 | 0.5450 | 0.3542 | 0.9273 |
| | November | 0.3579 | 0.4758 | 0.3044 | 0.9494 |
| | December | 0.2298 | 0.2094 | 0.1344 | 0.9114 |
| | **Average** | 0.3993 | 0.4728 | 0.2999 | 0.9298 |

**Table S7.**

**Leaderboard of normalized annual precipitation in China's five major river basins during the Ming and Qing dynasties.** The table presents the top three flood and drought events ranked by normalized precipitation magnitude for each river basin, including the event year, averaged normalized precipitation, and the standard deviation of the ensemble. Positive/negative values indicate floods/droughts, with larger absolute values representing greater intensity.

| River Basin | Category | Year | Averaged normalized annual precipitation | Standard deviation of ensemble |
|---|---|---|---|---|
| Hai River | Flood | 1801 | 2.87 | 0.15 |
| | | 1822 | 2.86 | 0.16 |
| | | 1762 | 2.25 | 0.17 |
| | Drought | 1609 | -2.44 | 0.09 |
| | | 1877 | -2.25 | 0.10 |
| | | 1689 | -2.22 | 0.10 |
| Yellow River | Flood | 1822 | 2.26 | 0.18 |
| | | 1662 | 2.26 | 0.18 |
| | | 1593 | 2.15 | 0.22 |
| | Drought | 1877 | -3.12 | 0.07 |
| | | 1720 | -2.60 | 0.09 |
| | | 1900 | -2.36 | 0.10 |
| Huai River | Flood | 1593 | 3.33 | 0.12 |
| | | 1746 | 2.11 | 0.14 |
| | | 1709 | 2.01 | 0.15 |
| | Drought | 1785 | -2.56 | 0.08 |
| | | 1640 | -2.39 | 0.08 |
| | | 1856 | -2.33 | 0.10 |
| Yangtze River | Flood | 1682 | 1.52 | 0.13 |
| | | 1832 | 1.43 | 0.14 |
| | | 1831 | 1.40 | 0.13 |
| | Drought | 1778 | -1.97 | 0.11 |
| | | 1900 | -1.71 | 0.10 |
| | | 1528 | -1.62 | 0.09 |
| Pearl River | Flood | 1713 | 2.33 | 0.21 |
| | | 1633 | 1.96 | 0.29 |
| | | 1571 | 1.92 | 0.22 |
| | Drought | 1902 | -2.35 | 0.15 |
| | | 1721 | -2.05 | 0.18 |
| | | 1895 | -1.97 | 0.14 |

**Table S8.**

**Seasonal and annual averaged normalized precipitation statistics for China's five major river basins in 1593.** Positive/negative values indicate above-/below-normal precipitation, with larger absolute values representing greater deviations from the climatological mean. The values ranking top 3 for flood events during the Ming and Qing dynasties are marked with superscript numbers, which indicate their respective ranks.

| River basin   | Spring      | Summer      | Autumn      | Winter | Annual      |
|---------------|-------------|-------------|-------------|--------|-------------|
| Hai River     | 1.99        | 0.43        | 0.68        | -0.58  | 1.03        |
| Yellow River  | $3.03^{1}$  | 1.35        | 1.07        | -0.79  | $2.15^{3}$  |
| Huai River    | $3.32^{1}$  | $2.19^{1}$  | $2.03^{1}$  | -0.86  | $3.33^{1}$  |
| Yangtze River | -0.17       | 0.62        | 0.78        | -0.65  | 0.50        |
| Pearl River   | -0.40       | 0.01        | 0.00        | 0.63   | -0.21       |

**Table S9.**

**Seasonal and annual averaged normalized precipitation statistics for China's five major river basins in 1640.** Positive/negative values indicate above-/below-normal precipitation, with larger absolute values representing greater deviations from the climatological mean. The values ranking top 3 for drought events during the Ming and Qing dynasties are marked with superscript numbers, which indicate their respective ranks.

| River basin | Spring | Summer | Autumn | Winter | Annual |
|---|---|---|---|---|---|
| Hai River | -1.75 | $-1.95^{3}$ | -0.77 | 0.34 | -2.04 |
| Yellow River | -1.84 | $-2.06^{2}$ | -0.59 | 0.14 | -1.99 |
| Huai River | -1.18 | -1.71 | -1.73 | -0.84 | $-2.39^{2}$ |
| Yangtze River | -0.13 | -0.25 | -0.18 | -1.08 | -0.62 |
| Pearl River | -0.59 | 0.29 | 0.60 | -0.75 | -0.07 |

**Table S10.**

**Seasonal and annual averaged normalized precipitation statistics for China's five major river basins in 1877.** Positive/negative values indicate above-/below-normal precipitation, with larger absolute values representing greater deviations from the climatological mean. The values ranking top 3 drought events during the Ming and Qing dynasties are marked with superscript numbers, which indicate their respective ranks.

| River basin | Spring | Summer | Autumn | Winter | Annual |
|---|---|---|---|---|---|
| Hai River | -0.88 | $-2.02^{1}$ | -1.37 | -1.29 | $-2.25^{2}$ |
| Yellow River | -1.80 | $-2.12^{1}$ | $-2.19^{1}$ | $-2.05^{3}$ | $-3.12^{1}$ |
| Huai River | -1.49 | -1.41 | -0.69 | -0.97 | -1.85 |
| Yangtze River | 0.07 | -0.55 | -1.14 | $-2.19^{3}$ | -1.03 |
| Pearl River | 0.78 | 1.29 | -0.57 | -1.70 | 0.84 |